\documentclass[aps,prb,reprint,superscriptaddress,amsmath,amssymb]{revtex4-2}

\usepackage{graphicx}
\usepackage{bm}
\usepackage{hyperref}
\usepackage{bbm}
\usepackage{amsmath}
\numberwithin{equation}{section}
\usepackage{amssymb}
\newcommand{\Tr}{\mbox{Tr}}
\newcommand{\e}{\text{e}}

\begin{document}

\title{Length-resolved Operator Growth\\ and Path-Entropy Obstructions to Many-Body Localization}

\author{J.~Sirker}
\affiliation{Department of Physics and Astronomy,
University of Manitoba, Winnipeg, Canada R3T 2N2}

\date{\today}

\begin{abstract}
For the disordered Ising chain with transverse and longitudinal fields, where couplings and fields are drawn from strictly positive distributions, Cao~\cite{Cao} has shown that the moments $\mu_{2k} = \|[H,\sigma^z_0]^{(k)}\|_2^2$ grow almost factorially, $\mu_{2k}^{1/(2k)}\sim k/\ln k$, and thus asymptotically at the maximal allowed rate. We generalize this result by resolving the operator norm in support length and show that the weight at length $\ell_k \sim k/\ln k$ already exhibits almost factorial growth, $\|[H,\sigma^z_0]^{(k)}_{\ell_k}\|_2 \gtrsim (k/\ln k)^k$. This implies maximal spatial delocalization of local operators and, in particular, rules out dynamical locality---the strongest form of many-body localization---at any disorder strength. We further establish rigorously a finite-size crossover scale $L\sim (W/J)^2$, where $W$ is the disorder and $J$ the coupling strength. For $L\lesssim (W/J)^2$ numerical studies only access a pre-asymptotic regime. Finally, we identify a structural path-entropy obstruction to perturbative LIOM constructions, based on the almost factorial branching of operator content and independent of resonance effects; the same mechanism strongly suggests ballistic real-time operator spreading, so sub-ballistic or localized dynamics would require a presently unidentified cancellation principle acting on almost factorially many disorder-dependent paths with random amplitudes.
\end{abstract}

\maketitle

\section{Introduction}
\label{Intro}
It is well known and has been proven rigorously that in one-dimensional non-interacting quantum systems with short-range hoppings, any amount of disorder leads to localization \cite{Anderson58,AbrahamsAnderson,EdwardsThouless,BucajDamanik}. This so-called Anderson localization is a wave phenomenon and as such can also occur in classical systems \cite{Page2008}. From a quantum perspective, the most interesting question is if localization survives if interactions between the particles are included. Importantly, many-body localization (MBL) cannot be understood as the localization of a wave function and requires alternative frameworks based on many-body observables. The question whether MBL exists has been the subject of controversial discussions in the last 15 years, both regarding its precise definition and its stability in the thermodynamic limit \cite{BaskoAleiner,OganesyanHuse,PalHuse,ZnidaricProsen,BardarsonPollmann,SchreiberHodgman,NandkishoreHuse,AltmanVoskReview,WeisseGerstnerSirker,SelsPolkovnikov,SelsPolkovnikovPRX,AbaninRev2019,KieferUnanyan2,KieferUnanyan3,SuntajsBonca,DeRoeckQuasilocalization}. One of the main issues in studying MBL is that there is no single definition and that aspects such as vanishing transport, retained memory, entanglement growth, level statistics, and emerging conservation laws are not necessarily all equivalent to each other. In practice, studies have often concentrated on exact diagonalizations of the Heisenberg chain with random magnetic fields, obtaining the properties of eigenspectra and analyzing quench dynamics \cite{ZnidaricProsen,OganesyanHuse,Luitz1,Luitz2}. However, the accessible system sizes are very limited making it difficult to distinguish a true phase transition from a finite-size crossover \cite{SuntajsBonca,SierantDelande,AbaninBardarson,SelsPolkovnikovPRX}. 

The perhaps most fundamental way of trying to define an MBL phase is to demand that in such a phase local integrals of motion (LIOMs) have to emerge. Furthermore, the microscopic Hamiltonian should be related to the LIOM Hamiltonian by a quasi-local unitary transformation. The latter is understood as a transformation that keeps local operators local up to exponential tails. It is also typically assumed that starting from strong disorder, the quasi-local unitary and therefore the LIOM Hamiltonian itself can be constructed perturbatively \cite{SerbynPapic,RosMuellerScardicchio,ChandranKim,HuseNandkishore,ImbrieRosScardicchio,Imbrie_JSTAT}. 

Fairly recently, a novel approach of studying ergodicity and ergodicity breaking has been put forward that is based on studying operator growth under Hamiltonian dynamics. It has been proven using combinatorial arguments that in a nearest-neighbor, one-dimensional model the maximal possible growth rate is almost factorial \cite{AvdoshkinDymarsky} and it has been hypothesized that every ergodic system will asymptotically grow consistently with this maximal rate \cite{ParkerCao,HevelingWang}. For the Ising chain with random transverse and longitudinal fields, where the couplings are drawn from strictly positive distributions, Cao \cite{Cao} has, furthermore, rigorously proven that the operator growth also obeys this almost factorial rate asymptotically. In addition, evidence from symbolic calculations strongly suggests that the same is also true in the Heisenberg model with random magnetic fields \cite{WeisseGerstnerSirker}. This raises a structural tension: if maximal operator growth persists in interacting disordered systems, it is unclear whether any meaningful notion of spatial locality can be maintained in the thermodynamic limit.

The main conclusion of our paper is that maximal operator growth leaves very little room for the usual localized-operator picture: dynamical locality is rigorously ruled out, real-time localization would require systematic cancellations of almost factorially many disorder-dependent paths (which we refer to as path entropy in the following) with random amplitudes, and perturbative LIOM constructions face a path-entropy obstruction distinct from resonances and avalanches \cite{DeRoeckHuveneers,DeRoeckHuveneers2,ThierryHuveneers,CrowleyChandran,MorningstarColmenarez}. The crucial point of our study is that rigorous results are not limited to the asymptotic scaling of moments $\mu_{2k}$ \cite{Cao} but that bounds on the spatial distribution of operator weight---which is directly related to questions of locality---can be rigorously established as well. One immediate consequence of this {\it local bound} on operator weight is a crossover scale $L\sim (W/J)^2$ in a finite system of size $L$: For $L\lesssim (W/J)^2$, where $W$ is the disorder and $J$ the coupling scale, the system is in a pre-asymptotic regime. This severely limits the usefulness of numerical methods to study the stability of MBL at strong disorder $W/J$. This conclusion is consistent with the physical picture of Sels and
Polkovnikov \cite{SelsPolkovnikovPRX}, who argued that nonergodic behavior at strong disorder is a large but ultimately transient crossover regime. The mechanism discussed here is quite different though: instead of impurity relaxation, we identify a path-entropy obstruction in operator space.

As a first step in substantiating these conclusions, we derive a strict lower bound on local operator growth which establishes almost factorial weight locally at a length scale which grows approximately linearly, up to logarithmic corrections, with the commutator order. Based on this strong delocalization of operator weight, we study the resulting structural obstructions to MBL. We concentrate on the random Ising model with couplings drawn from strictly positive distributions which allows us to establish strict positivity of path contributions in the operator-growth bound. We note that this includes the case studied by Imbrie~\cite{Imbrie_JSTAT}, where an iterative construction of LIOMs was proposed to converge under certain assumptions. We do not analyze that construction here; instead, we identify a structural obstruction that applies to any perturbative scheme and then discuss what would be required to overcome this obstruction. Our results fall into two categories: (1) Rigorous results for the local distribution of operator weight which imply that local operators do not stay local under Hamiltonian dynamics thus excluding dynamical locality. These results also imply a rigorous finite-size crossover scale $L\sim (W/J)^2$, see above. (2) Implications of our local operator growth result for real-time dynamics and the existence of LIOMs. In particular, we analyze under which conditions a perturbative LIOMs construction fails and show that, independent of the problem of many-body resonances, the main issue is to control the almost factorially large branching of operator content. To avoid misunderstandings, we stress right from the start that the obstruction is not the $4^\ell$ different Pauli strings that exist for a given support length $\ell$. This exponential growth in the number of terms with support length could, in principle, be overcome by sufficiently small amplitudes decaying as $\e^{-\gamma\ell}$ for each of these strings. Instead, the obstruction has its origin in the large number of commutator paths that result in the same Pauli string.

Our paper is organized as follows: In Sec.~\ref{Model}, we review the basic operator formalism and define the Ising model. In Sec.~\ref{Cao_bounds}, we derive a lower almost factorial bound for the local operator growth. This proof uses a gauge transformation for the Ising model to obtain strictly positive path contributions, generalizing and simplifying the approach first discussed by Cao \cite{Cao}. We show in Sec.~\ref{Sec_OG_gen} that such an operator growth quite generally means that almost all the weight is caused by histories which spread and that the contribution from histories that remain inside a fixed finite region is asymptotically negligible. In Sec.~\ref{Dyn_loc}, we show that this local growth is inconsistent with dynamical locality and derive a finite-size crossover scale. However, the operator growth does not refute abstract LIOMs in general and we argue in Sec.~\ref{LIOMs} that this is because quasi-locality of the unitary is too weak a constraint when dealing with almost factorial weights. We note that this is a statement about abstract LIOM Hamiltonians and certain quasi-local operators, not about whether or not a microscopic spin model can be mapped by a quasi-local unitary onto a specific LIOM Hamiltonian. The latter question is addressed in Sec.~\ref{Obstructions} where we argue that the path entropy due to almost factorial branching poses strong structural obstructions on a perturbative construction of LIOMs. In the final section \ref{Concl}, we summarize our main results and discuss the remaining open questions. 

\section{Definitions and model}
\label{Model}
We start by introducing the operators and norms studied, followed by a definition of the concrete model we consider when deriving rigorous results.

\subsection{Pauli strings and norms}
We study localization in spin chains made up of spin-$1/2$ operators on sites $j=1,\cdots,L$. We use the standard Pauli operators $\sigma^{x,y,z}_j$ plus the identity $\sigma^0_j\equiv\mathbbm{1}_j$. A Pauli string is then a tensor product
\begin{equation}
    \label{string}
s=\sigma_1^{\alpha_1}\otimes\sigma_2^{\alpha_2}\otimes\cdots\otimes\sigma_L^{\alpha_L}\quad\mbox{with}\quad \alpha_j\in\{0,x,y,z\}.
\end{equation}
We define the support of the Pauli string $s$ as the smallest contiguous interval $[j_\text{min}(s),j_\text{max}(s)]$ that contains every site $j$ with $\alpha_j\neq 0$. The length of the support is then given by
\begin{equation}
    \label{lengh}
    l(s)=j_\text{max}-j_\text{min}+1
\end{equation}
if $s$ is non-trivial, and $l(s)\equiv 0$ for the identity string. The Pauli strings form an orthonormal basis of operator space, $(s|s')=\delta_{s,s'}$, under the inner product
\begin{equation}
    \label{inner}
    (A|B) = \frac{\Tr(A^\dagger B)}{\Tr(\mathbbm{1})}
\end{equation}
for two operators $A,B$. Any operator can be expanded in the Pauli basis and written as
\begin{equation}
    \label{expansion}
    A=\sum_s a_s s \quad\mbox{with}\quad a_s=(s|A) \, .
\end{equation}
The normalized inner product \eqref{inner} induces the $2$-norm (Frobenius norm)
\begin{equation}
    \label{norm}
    \| A \|_2 = \sqrt{(A|A)}=\sqrt{\sum_s |a_s|^2}\, .
\end{equation}
The $2$-norm satisfies $\|s\|_2=1$ for any Pauli string including the identity and, importantly for what follows, is invariant under unitary transformations $U$,
\begin{equation}
    \label{unitary_cond}
    \|UAU^\dagger\|_2=\|A\|_2\, ,
\end{equation}
which is a direct consequence of the cyclic invariance of the trace. The normalized $2$-norm is sub-additive, $\|A+B\|_2\leq \|A\|_2 +\|B\|_2$ but, due to the normalization factor $\Tr(\mathbbm{1})$, is not sub-multiplicative. However, for the special case where $A=J_A s_A$ and $B=J_B s_B$ (both operators are individual Pauli strings, not sums) we have $\|AB\|_2=\|A\|_2\|B\|_2=|J_AJ_B|$.

We can resolve norms by the support lengths of their operator content by defining
\begin{equation}
    \label{l_resolution}
    A=\sum_\ell A_\ell,\quad A_\ell=\sum_{s,\, \ell(s)=\ell} a_s s,\quad \|A\|^2_2=\sum_\ell \|A_\ell\|_2^2 \, .
\end{equation}
This rearrangement of the sum is exact because of the orthogonality property of the Pauli strings and the fact that $A_\ell$ for different $\ell$ contain strings from disjoint sets of the Pauli basis.

\subsection{Iterated commutators and moments}
One of the central objects of our study is the $k$-fold commutator of a spin-chain Hamiltonian $H$ with an operator $A$ which we define as
\begin{equation}
\label{comm}
    \mathcal{L}_H^k(A)=[H,A]^{(k)}=[H,[H,\cdots,[H,A]]]\, .
\end{equation}
We will consider length-resolved norms of this iterated commutator given by
\begin{eqnarray}
    \label{comm2}
    s^{\ell,k}_H(A)&=&\|[H,A]_\ell^{(k)}\|_2\quad\mbox{and} \nonumber \\
    s_H^k(A)&=&\|[H,A]^{(k)}\|_2=\sqrt{\sum_\ell \left(s^{\ell,k}_H(A)\right)^2} \, .
\end{eqnarray}
These quantities are directly related to the moments of the infinite-temperature spectral function associated with $A$ by the relation
\begin{equation}
    \label{moment}
    \mu_{2k}=(A|\mathcal{L}_H^{2k}(A))=\|[H,A]^{(k)}\|_2^2=(s^k_H(A))^2\geq 0 \, .
\end{equation}
Note that the Liouvillian $\mathcal{L}_H$ is Hermitian with respect to the inner product \eqref{inner}.

\subsection{Quasi-locality}
\label{Locality}
There are different notions of quasi-locality and clearly distinguishing between them is important for the main results in this paper. If we have an operator $A$, then we can expand it in the Pauli string basis, $A=\sum_s a_s s$. A coefficient-based version of quasi-locality then is to demand that there exist a constant $C_A>0$ and a site $j_0$ at which the operator $A$ is centered at such that
\begin{equation}
    \label{local}
    |a_s|\leq C'_A \text{e}^{-\kappa_1\, d(s,j_0)}\text{e}^{-\kappa_2 \ell(s)}
\end{equation}
where $d(s,j_0)=\mbox{dist}(j_0,[j_\text{min}(s),j_\text{max}(s)])$ is the distance from $j_0$ to the support of $s$ and $\kappa_{1,2}>0$ are constants. This notion of locality is sufficient in the Anderson case where we are only dealing with one-body operators and the number of allowed Pauli strings with length $\ell$ is always $\mathcal{O}(1)$ independent of $\ell$. In the many-body case, on the other hand, all types of strings are allowed in general and the number of Pauli strings of length $\ell$ scales as $4^\ell$. This already shows that coefficient-wise exponential decay is not equivalent to aggregate quasi-locality. In the iterated commutators studied below, an additional path entropy arises from the many commutator histories contributing to the same final Pauli strings, and this can overwhelm the decay of individual path amplitudes.

The relevant notion of quasi-locality in a many-body system is therefore aggregate and norm based. In general, we can define
\begin{equation}
    \label{local2}
    A_{d,\ell}=\sum_{\substack{s,d(s,j_0)=d, \\ \ell(s)=\ell}} a_s s
\end{equation}
and demand that
\begin{equation}
    \label{local2_2}
    \|A_{d,\ell}\|_2\leq C_A \,\text{e}^{-\gamma_1 d}\text{e}^{-\gamma_2 \ell}
\end{equation}
with $C_A,\gamma_{1,2}>0$ constant. The norm-based definition \eqref{local2} implies the coefficient-based definition \eqref{local} because for each coefficient $a_s$ of a string $s$ with length $\ell$ at distance $d$ we have
\begin{equation}
    \label{local3}
    |a_s|\leq \sqrt{\sum_{\substack{s,d(s,j_0)=d,\\ \ell(s)=\ell}}|a_s|^2}=\|A_{d,\ell}\|_2\leq C_A \text{e}^{-\gamma_1 d}\text{e}^{-\gamma_2 \ell}\, .
\end{equation}
However, the converse is not true because there are exponentially many possible Pauli strings of length $\ell$; in the dynamical setting below, this static operator entropy is further amplified by a commutator path entropy and this entropy can overwhelm the exponential coefficient decay. We note that we can sum over distances $d$ to obtain the bound
\begin{equation}
    \label{local4}
    \|A_\ell\|_2\leq\sum_d \|A_{d,\ell}\|_2 \leq C_A' \text{e}^{-\gamma_2 \ell}
\end{equation}
which is the main aggregate definition of quasi-locality that we will use in the following.

\subsection{Dynamical Locality}
\label{dyn_loc}
The strongest version of MBL is to demand that---just as in Anderson localization---every initially local observable becomes at most quasi-local under Hermitian dynamics. More precisely, we demand that constants $C_A,\Lambda_A,\gamma_A>0$ exist such that for every local observable $A$ we have
\begin{equation}
    \label{dyn_loc1}
s_H^{\ell,k}(A)=\|[H,A]_\ell^{(k)}\|_2\leq C_A \Lambda_A^k \text{e}^{-\gamma_A \ell} \, .
\end{equation}
The constants can depend on the operator but not on the commutator order $k$ or the support length $\ell$. The factor $\Lambda_A^k$ allows for the overall norm of the commutator to grow exponentially with $k$ but this growth is always accompanied by an exponential tail which remains unchanged. We will show that this strong notion of locality is in contradiction with the exact local bounds for norm growth in the Ising chain which we derive later. In the Anderson case, on the other hand, dynamical locality can be proven rigorously \cite{WeisseGerstnerSirker}.  

\subsection{Effective LIOM Hamiltonian}
A LIOM Hamiltonian is defined by
\begin{equation}
    \label{LIOMH}
    \tilde H = \sum_n E_n \tau^z_n +\sum_{n_1<n_2} J_{n_1n_2}\tau^z_{n_1}\tau^z_{n_2}+\cdots
\end{equation}
where the $\tau^z_n$ are all mutually commuting and also commuting with the Hamiltonian. For the exchange coefficients we demand that $|J_{n_1\cdots n_r}|\leq J_0\exp(-\alpha(n_r-n_1))$ with $\alpha>\ln 2$. I.e., the decay rate $\alpha$ exponentially suppresses multi-spin interactions and ensures that $\tilde H$ is well-defined and its energy density convergent. One can construct operators $\tau_n^{x,y}$ which, together with $\tau_n^z$, fulfill the usual spin algebra. These operators generate another orthonormal Pauli-string basis and every operator can be expanded in this basis as well.

\subsection{LIOM-MBL definition}
The perhaps most fundamental way to define MBL is to demand that quasi-local LIOMs have to exist and that the microscopic Hamiltonian is unitarily equivalent to a LIOM Hamiltonian.

More precisely, we demand that there exists a unitary $U$ fulfilling the following two conditions:
\begin{itemize}
    \item[(1)] $UHU^\dagger=\tilde H$, with $\tilde H$ being of the LIOM form \eqref{LIOMH}.
    \item[(2)] Every strictly local operator $A$ in the microscopic $\sigma^{x,y,z}$ basis is transformed by the unitary transformation $U$ into a dressed operator $\tilde A=UAU^\dagger$ which is quasi-local, as defined in Eq.~\eqref{local4}, in the $\sigma^{x,y,z}$ basis. 
\end{itemize}
The last condition implies that spatial locality is preserved and $U$ itself is quasi-local. Without it, $U$ could be completely non-local and the transformation would describe a general diagonalization of the Hamiltonian. As has already been shown in Ref.~\cite{WeisseGerstnerSirker}, this definition of MBL is weaker than the dynamical locality defined in Sec.~\ref{dyn_loc}. Dynamical locality does imply the LIOM-MBL definition but the opposite is not true because both $\tilde H$ and $\tilde A$ are quasi-local and therefore $\|[\tilde H,\tilde A]^{(k)}\|_2=\|[H,A]^{(k)}\|_2$ is not forbidden by the LIOM-MBL definition to grow at the maximal, almost factorial rate. However, we will see later that there are strong structural obstructions against a perturbative construction of LIOMs in models with almost factorial operator growth.

\subsection{The Ising model with random couplings}
In the following, we concentrate on the one-dimensional random Ising model with longitudinal and transverse fields
\begin{equation}
    \label{Ising}
    H=\sum_j\left[J_j \sigma^z_j\sigma^z_{j+1}+h_j^x\sigma^x_j+h_j^z\sigma^z_j\right]
\end{equation}
when deriving rigorous results. We demand that the couplings satisfy two-sided bounds
\begin{equation}
    \label{cond}
    J_- \leq |J_j| \leq J_+,\quad h^x_- \leq |h^x_j| \leq h^x_+,\quad h^z_- \leq |h^z_j| \leq h^z_+
\end{equation}
with $0<J_-\leq J_+$, $0<h^x_-\leq h^x_+$, and $0<h^z_-\leq h^z_+$.

The cases where we can derive exact lower bounds for operator growth are those where $J_j>0$ and $\varepsilon_x,\varepsilon_z\in \{-1,+1\}$ exist such that $\varepsilon_x h^x_j>0$, and $\varepsilon_z h^z_j>0$ for all $j$. Then, we can always perform unitary transformations with either $\prod_j \sigma^z_j$ or $\prod_j\sigma^x_j$ or a combination of both to transform the Hamiltonian into a Hamiltonian with $h^x_j,h^z_j>0$. We can therefore assume w.l.o.g.~that all the couplings are drawn from a strictly positive interval.

We remark that cases where e.g.~the magnetic fields are drawn from distributions without a definitive sign are related to those with a definitive sign by adding appropriate constant magnetic fields. While in these cases the techniques used to derive the rigorous lower bound no longer apply, it seems physically unlikely that random models which only differ by a constant magnetic field have an asymptotically different operator growth. We also note that in the Heisenberg chain with random fields, where the known techniques to derive a lower bound are also not applicable, symbolic calculations have shown an operator growth which is fully consistent with a maximal, almost factorial growth \cite{WeisseGerstnerSirker}. While our rigorous results are for the Ising model \eqref{Ising} with strictly positive couplings, we therefore hypothesize that all disordered nearest-neighbor spin chain models show asymptotically maximal, almost factorial, operator norm growth.

\section{Bounds on operator growth}
\label{Cao_bounds}
We start by briefly recalling the combinatorial arguments first given in Ref.~\cite{AvdoshkinDymarsky}, see also Ref.~\cite{WeisseGerstnerSirker}, which establish a general almost factorial upper bound for operator norm growth in one-dimensional nearest-neighbor models without repeating the entire derivation here. The central question, addressed thereafter, is if this maximal growth rate is actually realized and how the operator weight is distributed.

Starting point is a Hamiltonian $H=\sum_j h_j$, where $h_j$ acts on nearest-neighbor sites only, and a fully local operator $A$. If one considers now the commutator $[H,A]^{(k)}$ then the main realization is that at every step a connected cluster has to be formed by $A$ and the local Hamiltonians $h_j$ because otherwise the commutator vanishes. The contribution to the total norm of each one of these clusters can be bounded using the largest scale present in the microscopic Hamiltonian and the remaining task is to count all the possible connected clusters at commutator order $k$. This leads to the general upper bound
\begin{equation}
    \label{upper}
    \|[H,A]^{(k)}\|_2\leq \|A\|_2 \Lambda^k B_k(2)\sim (k/ \ln k)^k
\end{equation}
where $\Lambda$ is a constant depending on the microscopic parameters of the model and $B_k(2)$ is the Bell (Touchard) polynomial of order $k$. However, nothing guarantees that the operator norm grows at this maximal rate. In principle, there could be cancellations within the almost factorial branching of connected clusters which lead to an asymptotically slower growth. In our view, however, this appears unlikely in disordered systems because the couplings along each path are products of random amplitudes and such cancellations would require some precise fine-tuning between almost factorially-many different commutator paths. 

In the Ising model \eqref{Ising} with positive couplings, this question can be resolved rigorously because a gauge exists in which all path contributions are positive. This was first realized in Ref.~\cite{Cao} and makes it possible to derive also a rigorous lower bound which turns out to show the same asymptotic scaling as the upper bound \eqref{upper}. Both bounds show almost factorial growth asymptotically and only differ by their amplitudes. 

\subsection{Length-resolved lower bound}
\label{Cao_local}
Here we will directly derive a length-resolved lower bound. The lower bound for the total norm  then follows as a corollary. We note that following Cao \cite{Cao} and working directly with the total norm is useful to make the bound as tight as possible by fixing the prefactors of the asymptotic scaling. Here, however, our focus is on how the operator weight is distributed in support length, rather than on optimizing prefactors.

The first step is to introduce a new basis 
\begin{equation}
    \label{basis}
    \tilde\sigma^0=\sigma^0,\; \tilde\sigma^x=-\sigma^x,\; \tilde\sigma^y=-i\sigma^y,\;\tilde\sigma^z=\sigma^z 
\end{equation}
which leads to the transformation for the Pauli string $\tilde s_{\alpha} =\tilde\sigma_1^{\alpha_1}\otimes\cdots\otimes\tilde\sigma_L^{\alpha_L}=\phi(s_\alpha)s_\alpha$ with a phase factor
\begin{equation}
    \label{phase}
    \phi(s)=(-1)^{n_x(s)}(-i)^{n_y(s)}
\end{equation}
where $n_{x,y}(s)$ counts the number of $\sigma^{x,y}$ factors in $s$, respectively. We note that the new basis is not hermitian because $(\tilde\sigma^y)^\dagger=-\tilde\sigma^y$ is anti-hermitian. However, this is just a property of the basis and hermitian operators expanded in this basis will still remain hermitian. The new operators fulfill the commutator algebra
\begin{equation}
    \label{algebra}
    [\tilde\sigma^z,\tilde\sigma^x]=+2\tilde\sigma^y,\; [\tilde\sigma^z,\tilde\sigma^y]=+2\tilde\sigma^x,\; [\tilde\sigma^x,\tilde\sigma^y]=-2\tilde\sigma^z \, .
\end{equation}
The imaginary unit has been absorbed and the only negative coefficient occurs in the third relation. The Hamiltonian of the Ising model in the $\tilde\sigma^{x,y,z}$ basis looks the same as Eq.~\eqref{Ising} with the only modification being the replacement $h^x_j\to -h_j^x$.

{\bf Theorem 1:} In this gauge, all matrix elements produced by $\mathcal{L}_H$, where $H$ is the gauged Hamiltonian, are positive real numbers which are either $2J_j,2h^x_j$, or $2h^z_j$ if $J_j,h^x_j,h^z_j>0$.

{\bf Proof:} We have to consider the action of the three parts of the gauged Ising Hamiltonian on a general Pauli string $\tilde s$. The longitudinal field $h^z_j\tilde\sigma^z_j$ acts non-trivially on Pauli strings containing a $\tilde\sigma^{x,y}_j$ which gets flipped to $\tilde\sigma^{y,x}_j$ with amplitude $2h^z_j$. Similarly, the transverse field $-h^x_j\tilde\sigma^x_j$ acts non-trivially on strings $\tilde s$ containing a $\tilde\sigma^{z,y}_j$ which gets flipped to $\tilde\sigma^{y,z}_j$ with amplitude $2h^x_j$ which follows again from the relations \eqref{algebra}. Note that the minus sign in the field cancels the minus sign in the last commutator relation in Eq.~\eqref{algebra}. Finally, for the bond term we have to consider the commutator
\begin{eqnarray}
 \label{bond}
&& J_j [\tilde\sigma^z_j\tilde\sigma^z_{j+1},\tilde\sigma^\alpha_j\tilde\sigma^{\alpha'}_{j+1}] \\
&=& J_j\left(\tilde\sigma^z_j \tilde\sigma^\alpha_j[\tilde\sigma^z_{j+1},\tilde\sigma^{\alpha'}_{j+1}]+[\tilde\sigma^z_j,\tilde\sigma^\alpha_j]\tilde\sigma^{\alpha'}_{j+1}\tilde\sigma^z_{j+1}\right). \nonumber
\end{eqnarray}
The bond commutator is non-zero only if the Pauli string anticommutes with exactly one of the two local $\tilde\sigma^z$ factors. Equivalently, exactly one of the two local factors on sites $j,j+1$ must be in $\{x,y\}$, while the other is in $\{0,z\}$. In this case, the commutator relations \eqref{algebra} show that the resulting matrix element is $+2J_j$. If both local factors are in $\{x,y\}$, the two terms in Eq.~\eqref{bond} produce the same output string with opposite signs and cancel.

{\bf Theorem 2:} For the Ising model \eqref{Ising} with $J_j,h^x_j,h^z_j>0$ and $A=\sigma^z_0$, there exists a length $\ell_k$ and a constant $c>0$ such that
\begin{equation}
    \label{bound}
    \|(\mathcal{L}^k_H A)_{\ell_k+1}\|_2\geq \left(c\frac{k}{\ln k}\right)^k \, .
\end{equation}

{\bf Proof:} We work in the basis $\tilde\sigma^{x,y,z}$ where $\mathcal{L}_H$ has only positive matrix elements according to theorem 1. For $\ell\leq\lfloor k/2\rfloor$ we construct a growth path of length $k$ starting from $\tilde\sigma^z_0$ to the string
\begin{equation}
    \label{xi_string}
    \xi_{\ell+1} = \tilde\sigma^x_0\otimes\tilde\sigma^x_1\otimes\cdots\otimes\tilde\sigma^x_{\ell-1}\tilde\sigma^z_\ell \, .
\end{equation}
One possible path is an alternating application of $-h_j\tilde\sigma^x_j$ and $J_j\tilde\sigma^z_j\tilde\sigma^z_{j+1}$. The first operator flips $\tilde\sigma^z_j\to\tilde\sigma^y_j$ while the second operator grows the string by one $\tilde\sigma^y_j\to\tilde\sigma^x_j\tilde\sigma^z_{j+1}$. Since all paths have positive amplitudes, we find that the growth path from $\tilde\sigma^z_0$ to $\xi_{\ell+1}$ has the lower bound
\begin{equation}
    \label{grow}
    C_\textrm{grow}(\ell+1)\geq (2h_-^x)^\ell (2J_-)^\ell \, .
\end{equation}
This path requires $2\ell$ applications of $\mathcal{L}_H$ leaving another $n=k-2\ell$ steps which we will use to keep the support length at $\ell+1$ and to scramble the string $\xi_{\ell+1}$ by applying the operator $h^z_j\tilde\sigma^z_j$. At each scramble step we can choose any of the $\ell$ sites with a $\tilde\sigma^x$ operator so there are $\ell^n$ such scramble paths with each of them contributing at least an amplitude $(2h^z_-)^n$. It is this large entropy coming from the almost factorial number of scramble paths which creates large operator weights. Concretely, we obtain the bound
\begin{eqnarray}
    \label{bound2}
    \|(\mathcal{L}_H^k\sigma^z_0)_{\ell+1}\|_2&\geq& \ell^{k-2\ell}(2h^z_-)^{k}\left(\frac{h^x_- J_-}{(h^z_-)^2}\right)^\ell \e^{-\mathcal{O}(\ell)} \nonumber\\
    &=& \ell^{-2\ell}(C\ell)^k \e^{-\gamma\ell}\e^{-\mathcal{O}(\ell)}
\end{eqnarray}
with $C=2h^z_-$ and $\gamma=\ln[(h^z_-)^2/(h^x_-J_-)]$. The factor $\e^{-\mathcal O(\ell)}$ accounts for the possibility that the positive path contributions are distributed over exponentially many final strings within the same support sector. As a final step in the proof, we need to optimize the length $\ell$ of the string $\xi_{\ell+1}$. The leading contribution comes from the entropy factor $\ell^{k-2\ell}$ counting the number of paths. Taking the logarithm and a derivative with respect to $\ell$, we have to solve
\begin{equation}
    \label{opt}
    \frac{k}{\ell_k}-2\ln\ell_k-2+\Gamma=0 \quad \Leftrightarrow\quad \ell_k=\frac{k}{2W(\frac{1}{2}\text{e}^{1-\Gamma/2}k)}
\end{equation}
where $\Gamma=-2\ln h^z_-+\ln h^x_- + \ln J_- $ is a constant and $W$ is the Lambert-W function. For $k\gg 1$ we can use the leading asymptotics of the Lambert-W function, $W(k)\sim\ln k$ and drop the constants inside the logarithm which leads to 
\begin{equation}
    \label{scale}
    \ell_k\sim \frac{k}{2\ln k} \, .
\end{equation}
Putting this back into Eq.~\eqref{bound2} we find
\begin{equation}
    \label{local_bound}
\|(\mathcal{L}_H^k\sigma^z_0)_{\ell_k+1}\|_2\geq \left(\frac{ck}{\ln k}\right)^k
\exp\!\left[\mathcal{O}\!\left(\frac{k\ln\ln k}{\ln k}\right)\right]
\end{equation}
which proves that there is almost factorial weight at the length scale $\ell_k$ given by Eq.~\eqref{scale}. In the following, we will set $\ell_k+1\sim\ell_k$ asymptotically.

{\bf Corollary:} The total norm grows almost factorially
\begin{equation}
    \label{total}
    \|\mathcal{L}_H^k\sigma^z_0\|_2 \sim \left(\frac{k}{\ln k}\right)^k \, .
\end{equation}
This follows immediately by combining the lower length-resolved bound \eqref{local_bound} with the general upper bound \eqref{upper}.

This theorem shows that the almost factorial growth of the total operator norm, which was already proven in Ref.~\cite{Cao}, cannot be attributed to a proliferation of short-range contributions. It necessarily involves operators whose support length diverges, implying a spatial delocalization. 

We end this section with a couple of remarks. First, if we consider richer models by, for example, adding a $\sigma^x_j\sigma^x_{j+1}$ term to the Hamiltonian, then the commutator $[\tilde\sigma^x_j,\tilde\sigma^y_j]=-2\tilde\sigma^z_j$ will be present without a compensating sign in the prefactor and it is impossible to restore positivity. More generally, a factorized phase such as Eq.~\eqref{phase} cannot lead to positivity in models with several non-commuting bond terms because the equations for the factorized gauge then become over-determined. This includes the XXZ model studied symbolically in Ref.~\cite{WeisseGerstnerSirker}. In these cases, it is an open question if it is possible to prove an almost factorial lower bound on operator growth by either using a more general gauge ansatz or by using an entirely different approach which does not rely on the positivity of every single operator growth path.

\section{Operator growth and delocalization}
\label{Sec_OG_gen}
Based on what we have learned rigorously about the Ising model, we can draw some general conclusions on models with almost factorial norm growth. In all such models, we expect a competition between exponential localization due to the growth path as in Anderson models, and an entropic scrambling contribution, see Eq.~\eqref{bound2}. For all such models---and we hypothesize that every generic one-dimensional disordered many-body system with short-range couplings behaves this way---we can make the following general statement.

{\bf Theorem 3:} Consider a many-body system with $\|\mathcal{L}_H^k A\|_2\geq (ck/\ln k)^k$. Let $I$ be a finite interval containing the support of the operator $A$ and let $P_I$ project onto operators supported entirely within this interval. The contribution of paths which stay entirely inside of $I$ is $C^{(k)}=(P_I\mathcal{L}_H P_I)^k A$ and is bounded by $\|C^{(k)}\|_2\leq \Gamma_I^k \|A\|_2$ implying
\begin{equation}
    \label{Theorem3_1}
    \frac{\|C^{(k)}\|_2}{\|\mathcal{L}_H^k A\|_2}\leq \Gamma_I^k \|A\|_2 \left(\frac{\ln k}{ck}\right)^k \to 0\,.
\end{equation}
The almost factorial operator growth is generated by paths that leave any fixed region. Spreading is forced.

{\bf Proof:} Let $H=\sum_j h_j$ and let $O$ be an operator supported entirely within $I$. This implies that
\begin{eqnarray}
    \label{Theorem3_2}
    \|(P_I\mathcal{L}_H P_I) O\|_2&\leq&\sum_{j\in I} \|[h_j,O]\|_2 \\
    &\leq& 2 \sum_{j\in I} \|h_j\|_\infty \|O\|_2 \equiv \Gamma_I \|O\|_2\, . \nonumber
\end{eqnarray}
Iterating this equation immediately implies $\|(P_I\mathcal{L}_H P_I)^k A\|_2\leq \Gamma_I^k \|A\|_2$ with $\Gamma_I<\infty$ and Eq.~\eqref{Theorem3_1} follows. 

We note that this does not imply that $\|[H,A]^{(k)}_{\leq r_0}\|_2$ grows only exponentially, because this quantity also contains excursion-and-return paths that leave $I$ at intermediate steps. The theorem shows instead that any contribution growing faster than exponentially must arise from paths that leave every fixed finite region. More specifically, since $\Gamma_I$ grows only linearly with the size of the interval $I$, the same bound applied to windows of growing size $|I|=m$ shows that any contribution matching the almost factorial growth $(ck/\ln k)^k$ must involve paths exploring regions of size $m \gtrsim k/\ln k$, i.e., of the order of the saddle-point length $\ell_k$.

\section{Absence of dynamical locality and finite-size crossover}
\label{Dyn_loc}
For dynamical locality, the condition that local operators remain exponentially localized under Hamiltonian dynamics, we require that Eq.~\eqref{dyn_loc1} is fulfilled. Obviously, the lower local bound \eqref{local_bound} breaks this condition to the maximally allowed extent. The Ising model with positive couplings does not show dynamical locality. This appears to be in tension with exact diagonalizations of small systems that appear consistent with a localized dynamics. Here, it is important to stress that there is a competition between the exponential decay required for dynamical locality and the entropic contribution due to almost factorial branching. If we start from the limit $h^z_-\gg h^x_-,J_-$, where the $\sigma^z_j$ operators are close to being LIOMs, we see from Eq.~\eqref{bound2} that the last two terms on the r.h.s.~are exactly of the form \eqref{dyn_loc1} expected for dynamical locality. In particular, the last term describes exponential localization with increasing length of the cluster. However, this term does get overwhelmed eventually by the $\ell^{k-2\ell}$ factor that comes from the almost factorial branching of scramble paths. This results in the following finite-size crossover scale separating a pre-asymptotic from the asymptotic regime.\\

{\bf Theorem 4 (pre-asymptotic regime):} The length-resolved bound \eqref{bound2} for the Ising chain implies that for a system of size $L$ with accessible commutator order $k\lesssim L$ the path entropy contribution dominates the exponential decay once
\begin{equation}
    \label{condition}
    L\gg \left(\frac{W}{J}\right)^2 \, .
\end{equation}
Here, $W=h^z_-$ and $J=\sqrt{h^x_- J_-}$ are the disorder and coupling scale, respectively.\\

{\bf Corollary:} The apparent localization observed in numerical studies for system sizes $L\lesssim (W/J)^2$ addresses a pre-asymptotic regime. The asymptotic operator delocalization sets in at the latest for $L\sim (W/J)^2$.\\

{\bf Proof:} The bound \eqref{bound2} for the Ising chain shows that the path entropy is competing with an exponential localization factor. The path entropy wins over the exponential localization factor if $\ell^k \gg \e^{\gamma\ell}$, i.e., if $k\ln \ell\gg \gamma \ell$. In a finite system, $L$ acts as an infrared cutoff: operators cannot spread over more than $L$ sites, which imposes a hard upper bound on the growth of the scrambling path entropy. The asymptotic path-entropy-dominated regime can therefore only be reached once the largest available support $\ell\sim L$ is sufficient for the entropy to overcome the exponential decay. Taking $k\sim L$ and $\ell\sim L$ gives $L\gg \e^\gamma=(W/J)^2$.

Since we expect that the scaling \eqref{bound2} will hold for disordered many-body spin systems in general, we also expect that the crossover phenomenon \eqref{condition} is general. While such a crossover has been observed numerically before \cite{SuntajsBonca}, with the empirical scaling $L\sim W/J$
being faster than our rigorous {\it lower bound} $L\sim (W/J)^2$, our result provides a rigorous foundation for why a finite-size crossover occurs and why the crossover scale grows with disorder strength. Physically, a system with $L\ll (W/J)^2$ will be close to fulfilling the dynamical locality condition \eqref{dyn_loc1} with the path entropy term only leading to small corrections. In this regime, the system therefore displays Anderson-like quasi-local behavior on the accessible length scales.

So far, the results in this section have been rigorous but while the rigorous bound (\ref{local_bound}) rules out dynamical locality, it does not by itself determine the real-time spreading of operators. We will end this section by discussing what can be said about the real-time dynamics in a system with almost factorial operator norm growth. The Lieb-Robinson bound provides a rigorous upper bound on the rate at which operators spread in real time in a model with short-range exchange: $A(t)$ has support within the cone $\ell \le vt$ (with $v$ the Lieb-Robinson velocity determined by the local structure of $H$) \cite{LiebRobinson,BravyiHastings}, and weight outside this light cone is exponentially suppressed. Whether this upper bound is saturated---i.e., whether real-time spreading is ballistic---depends on the internal structure of $A(t)$ within the cone. For Anderson-localized non-interacting systems, the cone is not saturated: the euclidean operator norm is exponentially localized so that $A(t)$ can spread at most logarithmically \cite{WeisseGerstnerSirker}. Due to the violation of dynamical locality in the interacting case, this Anderson-type real-time localization is rigorously excluded asymptotically for the Ising chain with positive couplings. The remaining possible mechanism for sub-ballistic spreading would be destructive interference among paths inside the cone---random amplitudes conspiring to suppress weight at lengths $\ell\lesssim vt$. Such cancellations would have to act systematically across almost factorially-many paths with random amplitudes, and for almost every disorder realization. This is non-generic: for sums of independent random variables and in the absence of symmetries or conservation laws that would enforce structural cancellation, generic disordered systems typically do not produce it. We therefore expect, by combining rigorous results with physical expectations for a system with random disorder, that real-time operator spreading in this model in the asymptotic limit is ballistic up to logarithmic corrections, saturating the Lieb-Robinson cone.

\section{Operator growth and abstract LIOM Hamiltonians}
\label{LIOMs}
The almost-factorial operator growth established in the previous sections does not by itself refute the possibility that a quasi-local unitary exists that maps the microscopic Hamiltonian to a LIOM Hamiltonian. In this section we show that operator growth on either side of such a hypothetical mapping is consistent: by unitary invariance the growth rates must match (Theorem 5), and an explicit example shows that LIOM Hamiltonians in general can support almost-factorial growth on quasi-local operators (Theorem 6). These results establish {\it consistency}, not {\it implication}: they show that almost factorial operator growth alone cannot exclude the LIOM-MBL hypothesis, but they do not establish that the microscopic Ising chain is connected to a LIOM Hamiltonian by any quasi-local unitary. The question of whether such a unitary can exist is a separate question which is addressed in the next section.

Consistency is fundamentally a consequence of the weakness of an exponential quasi-locality condition when dealing with almost factorial weights. To see this, consider an operator $O_r$ whose support is $r$ sites. We define a unitary as being quasi-local if
\begin{equation}
    \label{unitary}
   \|(\tilde O_r)_\ell\|_2= \|(U O_r U^\dagger)_\ell \|_2 \leq C_U \text{e}^{-\alpha(\ell-r)}\|O_r\|_2 
\end{equation}
for $\ell\geq r$. Now assume the operator has weight $\|O_r\|_2\sim (k/\ln k)^k$. As a consequence, the transformed operator can have weight $\|\tilde O_r\|_2\sim (k/\ln k)^k \text{e}^{-k/\ln k}$ at length $\ell_k\sim k/\ln k$. We see that the exponential is only a next leading correction and
is of the same order as next-leading corrections in the local operator growth bound. We conclude that quasi-locality as defined in Eq.~\eqref{unitary} is too weak to prevent factorial weight shift over distances $\ell_k$. For many-body systems with factorial operator weights, an exponential quasi-locality condition does not constrain the weight distribution much better than a generic non-local unitary. We note that this is very different from the non-interacting (Anderson) case where the entropic factor $\ell^k$ is absent in the many-body bound and weight is concentrated in a few matrix elements in the operator basis. In that case, quasi-locality is a strong and physically meaningful constraint. We next show first that if a quasi-local unitary mapping the microscopic Hamiltonian to a LIOM Hamiltonian exists, then operator growth passes the consistency check (theorem 5) before also showing that for abstract LIOM Hamiltonians, quasi-local operators exist which show norm growth even faster than the almost factorial nearest-neighbor result $(k/\ln k)^k$ (theorem 6). Note that such models are, in general, not nearest-neighbor models and the considered operators are quasi-local not local so that the upper bound \eqref{upper} does not apply.

{\bf Theorem 5:} Assume that a unitary $U$ exists mapping a microscopic Hamiltonian $H$ with almost factorial norm growth of a local operator $A$ under Hamiltonian dynamics to a LIOM Hamiltonian $\tilde H=UHU^\dagger$. Then the $k$-fold commutator $[\tilde H, \tilde A]^{(k)}$ of the LIOM Hamiltonian $\tilde H$ with the {\it quasi-local} operator $\tilde A=UAU^\dagger$ will also show almost factorial norm growth, $\|[\tilde H, \tilde A]^{(k)}\|_2\sim (k/\ln k)^k$. 
Second, let $P_I$ be the projector onto operators supported entirely within a finite interval $I$. Then the contribution generated by $P_I\tilde A$ from paths which remain entirely inside I is defined as $\tilde C^{(k)}=(P_I\mathcal{L}_{\tilde H} P_I)^k P_I\tilde A$ and is bounded by
\begin{equation}
    \label{Theorem5_1}
    \|\tilde C^{(k)}\|_2\leq \tilde \Gamma^{k}_I \|P_I\tilde A\|_2
\end{equation}
with $\tilde \Gamma_I<\infty$, implying
\begin{equation}
\label{Theorem5_2}
\frac{\|\tilde C^{(k)}\|_2}{\|\mathcal{L}^k_{\tilde H}\tilde A\|_2}\leq\tilde\Gamma^k_I\|P_I\tilde A\|_2 \left(\frac{\ln k}{ck}\right)^k\to 0\, .
\end{equation}
Thus, also in the LIOM picture, the almost factorial total norm growth necessarily involves operator content that lives outside any fixed region, either through the quasi-local tails of $\tilde A$ or through commutator paths that leave the region. Locality is not restored.

{\bf Proof:} The first part of the theorem follows directly from unitary invariance of the $2$-norm. Using the already proven upper and lower bounds we obtain
\begin{equation}
    \label{LIOM_bound}
    \left(\frac{c_- k}{\ln k}\right)^k \leq \|[H,A]^{(k)}\|_2=\|[\tilde H,\tilde A]^{(k)} \|_2\leq \left(\frac{c_+ k}{\ln k}\right)^k
\end{equation}
with $\tilde A=U A U^\dagger$. The proof of the second part of the theorem is completely analogous to the proof of theorem 3 with the projected operator $P_I\tilde A$ replacing the local operator $A$.

However, the dynamics in an abstract LIOMs model is even less restricted than in a microscopic nearest-neighbor model. This can be seen in a simple example first discussed in Ref.~\cite{WeisseGerstnerSirker}. 

{\bf Theorem 6:} Consider the trivial LIOMs Hamiltonian $\tilde H=\sum_n \varepsilon_n \sigma^z_n$ and the quasi-local operator
\begin{equation}
    \label{sigmaz}
    \tilde\sigma^z_0 = \sqrt{1-\sum_{\ell\geq 0} \text{e}^{-2\kappa (\ell+1)}}\sigma^z_0 +\sum_{\ell\geq 0} \text{e}^{-\kappa (\ell+1)}\sigma^y_0\otimes\cdots\otimes\sigma^y_{\ell-1}\sigma^x_\ell
\end{equation}
with $\kappa>\frac{1}{2}\ln 2$. Because the strings appearing in Eq.~\eqref{sigmaz} mutually anticommute and each squares to the identity, this operator fulfills $(\tilde\sigma^z_0)^2=\mathbbm{1}$ and therefore $\|\tilde\sigma^z_0\|_2=1$, as required for a unitary dressing of a local Pauli operator. Then, asymptotically, the lower length-resolved bound
\begin{equation}
    \label{Theorem5}
    \|[\tilde H,\tilde\sigma^z_0]^{(k)}_{\ell^*+1}\|_2 \geq 2\left(\frac{2\varepsilon_- k}{\text{e}(\kappa+\ln 2)}\right)^k
\end{equation}
at length $\ell^*\sim k/(\kappa+\ln 2)$ holds for $k$ even with $|\varepsilon_n|\geq\varepsilon_->0$. The total weight therefore grows faster than the almost factorial nearest-neighbor result \eqref{upper}, $\|[\tilde H,\tilde\sigma^z_0]^{(k)}\|_2 \geq (Ck)^k$, with some constant $C$.

{\bf Proof:} In general, the matrix element $([\tilde H,\cdot]^{(k)})_{s',s}$ is a sum over all paths of length $k$ from Pauli string $s$ to $s'$, and individual path amplitudes can interfere. It follows that a single path usually does not provide a lower bound on the matrix element. The present case, however, is special: $\tilde H = \sum_n \varepsilon_n \sigma^z_n$ consists of mutually commuting on-site terms, so the amplitude along a path depends only on the number of flips at each site, not on the order in which they are flipped. All paths between fixed $s$ and $s'$ therefore share a common complex amplitude and add coherently with no cancellations. 

We can expand the commutator norm as $\|[\tilde H,\tilde\sigma^z_0]^{(k)}_\ell\|_2^2=\sum_{s,\ell(s)=\ell} |c_s^{(k)}|^2$ with $[\tilde H,\tilde\sigma^z_0]^{(k)}=\sum_s c^{(k)}_s s$. Restricting to closed paths returning to $s' = s$ then gives a rigorous lower bound from any subset of such paths. Now fix $\ell \ge 1$ and consider the Pauli string $S_\ell = \sigma^y_0 \otimes \cdots \otimes \sigma^y_{\ell-1} \otimes \sigma^x_\ell$ of length $\ell+1$ appearing in $\tilde\sigma^z_0$ with coefficient $\text{e}^{-\kappa (\ell+1)}$. Each application of $[\tilde H,\cdot]$ flips the spin at one of the $\ell+1$ sites between $\sigma^x$ and $\sigma^y$, contributing a factor $\pm 2i\varepsilon_j$. Two flips at the same site give $(2i\varepsilon_j)(-2i\varepsilon_j) = 4\varepsilon_j^2 > 0$, so any closed path returning to $S_\ell$ has each site flipped an even number of times $n_j$ and contributes the positive amplitude $\prod_j (2\varepsilon_j)^{n_j}$ independent of the order the spins were flipped. What remains is counting the number of closed paths with even length $k$, given by a multinomial coefficient. This leads to the following lower bound 
\begin{eqnarray}
    c_{S_\ell}^{(k)} &=& \text{e}^{-\kappa (\ell+1)}\underbrace{\sum_{\stackrel{n_0+\cdots+n_\ell=k}{n_j\;\textrm{even}}}\!\!\!\frac{k!}{n_0!\cdots n_\ell!}}_{N^k_{\ell+1}}\prod_j (2\varepsilon_j)^{n_j} \nonumber \\ &\geq& \text{e}^{-\kappa (\ell+1)}(2\varepsilon_-)^k N^k_{\ell+1}
\end{eqnarray}
where the multinomial gives the closed form expression
\begin{equation}
    \label{Nkl}
    N^k_{\ell+1} = \frac{1}{2^{\ell+1}}\sum_r \binom{\ell+1}{r}(\ell+1-2r)^k\geq (\ell+1)^k/2^\ell
\end{equation}
where we have bounded the sum of positive terms by the $r=0$ and $r=\ell+1$ contributions. This leads to the bound 
\begin{equation}
    \|[\tilde H,\tilde\sigma^z_0]^{(k)}_{\ell+1}\|_2 \ge |c_{S_\ell}^{(k)}| \ge \frac{\text{e}^{-\kappa (\ell+1)}(2(\ell+1)\varepsilon_-)^k}{2^{\ell}}\, 
\end{equation}
for $k$ even. Finally, we can optimize the bound over $\ell$ which gives the saddle point $\ell^*+1=k/(\kappa+\ln 2)$ and the lower bound \eqref{Theorem5} which completes the proof. We stress again that a faster than $(k/\ln k)^k$ growth does not violate the upper bound \eqref{upper} which is only valid for nearest-neighbor Hamiltonians $H$ and fully local operators $A$.

While we have picked the specific operator $\tilde\sigma^z_0$ in theorem 6 for convenience, we note that the theorem is quite general. Any quasi-local operator which has sufficient off-diagonal content at any length scale to allow $\sim \ell^k$ scrambling paths will show almost factorial operator growth. 

In summary, almost-factorial operator norm growth is consistent with abstract LIOM Hamiltonians but this {\it does not establish} a connection between the microscopic Ising chain and any specific LIOM Hamiltonian. Whether such a connection exists is a separate question, depending---in the conventional definition of MBL---on whether LIOMs can be constructed perturbatively from the strong-disorder limit. The next section identifies a structural mechanism that obstructs convergence of such a construction.

\section{A path-entropy obstruction to perturbative  LIOMs}
\label{Obstructions}
The standard definition of MBL posits that the microscopic Hamiltonian can be mapped unitarily to a LIOM Hamiltonian. Furthermore, it is typically assumed that the limit $\lambda=J/W\to0$, where $W$ is the disorder strength, is non-singular and that the microscopic $\sigma^z_j$ are perturbatively dressed into quasilocal conserved quantities $\tau_j=\mathcal{U}\sigma^z_j\mathcal{U}^\dagger$ by a common quasi-local unitary $\mathcal{U}$. Based on this definition of MBL, we therefore have
\begin{center}
\fbox{
  \begin{minipage}{0.8\columnwidth}
  \centering
  conventional MBL \\*[1.8pt] $\;\Longleftrightarrow\;$ $\tau_j=\sigma^z_j+\sum_{n\geq 1} \lambda^n \tau_j^{(n)}$ quasi-local.
  \end{minipage}
}
\end{center}
We note that this definition of a conventional MBL phase does not exclude a different type of localized phase that only exists for finite $\lambda$ and is not analytically connected to the atomic limit. LIOMs in such an alternative localized phase could potentially be constructed through a multiscale Kolmogorov-Arnold-Moser (KAM) renormalization scheme \cite{Imbrie2016,Imbrie_JSTAT,deRoeckGiacomin} but this would still require understanding how such a construction can avoid the path-entropy obstruction discussed next and, if the scheme is convergent in the thermodynamic limit, would also physically describe a very different phase from the one typically understood as MBL. We will return to this point when discussing the implications of the obstruction theorem which we state next.

\subsection{Obstruction theorem}
{\bf Theorem 7 (conditional):} The operators in the perturbative construction of the quasilocal charges $\tau_j$ satisfy the lower bound $\|\tau_{j,\ell}^{(n)}\|_2^2\geq N(n,\ell)^\eta C^{-2n}$ with $C>0$ where $\tau_{j,\ell}^{(n)}$ is the part of the operator at order $n$ that is supported on $\ell$ sites and $N(n,\ell)$ is the number of paths which itself is bounded from below by $N(n,\ell)\geq \mathcal{B}(\ell)^{n-c\ell}$ with $c\in\mathbbm{N}$. The convergence radius of the support-$\ell$ contribution therefore satisfies $R_\ell\leq C \mathcal{B}(\ell)^{-\eta/2}$. 

If the number of scrambling paths $\mathcal{B}(\ell)$ for one step in the perturbative construction of the quasilocal charge $\tau_j$ diverges, $\mathcal{B}(\ell)\to\infty$ for $\ell\to\infty$, and if $\eta>0$ then the convergence radius $R_\ell$ vanishes as $\ell\to\infty$. Consequently, the perturbative construction of quasilocal LIOMs fails.


{\bf Proof:} We use
\begin{eqnarray}
    \label{Theorem7_1}
    H&=&H_0+\lambda V,\quad \tau_j=\sum_{n\geq 0} \lambda^n \tau_j^{(n)},\quad \tau_j^{(0)}=\sigma^z_j \nonumber \\
    && [H,\tau_j] = 0 ,\quad \tau_j^\dagger=\tau_j,\quad \tau_j^2=\mathbbm{1} 
\end{eqnarray}
where $\tau_j^2=\mathbbm{1}$ fixes the diagonal freedom for the conserved charges. Resolving the condition $[H,\tau_j]=0$ order by order in $\lambda$, we obtain the recursion relation
\begin{equation}
    \label{Theorem7_2}
    [H_0,\tau_j^{(n)}]=-[V,\tau_j^{(n-1)}] \, .
\end{equation}
We can separate $\tau_j^{(n)}$ into a part which is diagonal in the $H_0$ basis and a part which is off-diagonal by writing $\tau_j^{(n)}=P_d \tau_j^{(n)}+Q\tau_j^{(n)}$ where $P_d$ projects onto the diagonal basis and $Q=1-P_d$. Since $[H_0,P_d\tau_j^{(n)}]=0$ we can write the recursion relation \eqref{Theorem7_2} also as
\begin{equation}
    \label{Theorem7_3}
    Q\tau_j^{(n)} = R\tau_j^{(n-1)}
\end{equation}
with $R=-\mathcal{L}_0^{-1}Q\mathcal{L}_V$, $\mathcal{L}_0=[H_0,\cdot]$, and $\mathcal{L}_V=[V,\cdot]$. We note that this is again a repeated commutator similar in structure to $[H,A]^{(k)}$ considered earlier. We therefore expect that this commutator also shows extensive branching and that almost all $4^\ell$ possible Pauli strings of support length $\ell$ are eventually realized. We note, however, that the conditional theorem does not require extensive branching, $\mathcal{B}(\ell)\sim\ell$, only that $\mathcal{B}(\ell)\to\infty$ for $\ell\to\infty$. Here, the commutator is dressed by a bare energy difference
\begin{equation}
    \label{Theorem7_4}
    \left(\tau_j^{(n)}\right)_{\alpha\beta}=-\frac{[V,\tau_j^{(n-1)}]_{\alpha\beta}}{E_\alpha-E_\beta},\; \alpha\neq \beta
\end{equation}
but this does not change the branching properties of the commutator. In addition to the off-diagonal recursion relation \eqref{Theorem7_3}, there is also a relation that follows from $\tau_j^2=\mathbbm{1}$ and fixes the diagonal part. At order $n=0$ the condition is fulfilled, because $(\sigma^z_j)^2=\mathbbm{1}$. At higher orders, we obtain 
\begin{equation}
    \label{Theorem7_5}
    \{\sigma^z_j,\tau_j^{(n)}\} +\sum_{m=1}^{n-1}\tau_j^{(m)}\tau_j^{(n-m)}=0,\; n\geq 1 \, .
\end{equation}
Projecting onto the diagonal part then yields the recursion relation
\begin{equation}
    \label{Theorem7_6}
    P_d\tau_j^{(n)} =-\frac{1}{2}\sigma^z_j P_d\sum_{m=1}^{n-1}\tau_j^{(m)}\tau_j^{(n-m)},\; n\geq 1\, .
\end{equation}
Thus, the two recursion relations \eqref{Theorem7_3} and \eqref{Theorem7_6} completely determine the perturbative quasilocal charges $\tau_j$.

Eq.~\eqref{Theorem7_3} is again an iterated commutator with the difference---as compared to the previously studied case of $[H,A]^{(k)}$---that the amplitudes contain energy denominators, see Eq.~\eqref{Theorem7_4}. Since the hopping in $V$ extends the support of the operator by at most one site, the support of $\tau^{(n)}$ extends up to length $\ell\sim n$ in general. There will therefore be growth paths for $n\gtrsim\ell$ which extend the support of $\tau^{(n)}$ up to length $\ell$ in $c\ell$ growth steps. Once this support length is reached, there are parts of the commutator of $V$ with the existing string that remain off-diagonal and within the interval $I_\ell$. We call the number of such scrambling paths available at each step $\mathcal{B}(\ell)$ and there are $n-c\ell$ scrambling steps. I.e., this scrambling mechanism exists at each step of the iterated commutator \eqref{Theorem7_3}. As a consequence, the number of paths is bounded by
\begin{equation}
    \label{Theorem7_7}
    N(n,\ell)\geq \mathcal{B}(\ell)^{n-c\ell}
\end{equation}
with $c\in\mathbbm{N}$. We assume the best case scenario for MBL where resonances play no role and the matrix elements of $\tau_{j,\ell}^{(n)}$---the operator $\tau_j^{(n)}$ with support length $\ell$--- for each path are of order $C^{-n}$ with $C>0$. We note that the exponential suppression factor $\lambda=J/W$ has already been separated out, see Eq.~\eqref{Theorem7_1}. The constant $C$ is $\mathcal{O}(1)$ and accounts for other path-related contributions to the amplitude. For the construction of the quasilocal charges this implies that
\begin{equation}
    \label{Theorem7_8}
    \|\tau_{j,\ell}^{(n)}\|_2^2\geq N(n,\ell)^\eta C^{-2n} \, 
\end{equation}
with $\eta$ describing how the paths add up. In particular, $\eta=2$ for a coherent addition of paths and $\eta=1$ in the incoherent case. The latter is what might be expected if the contributions from different paths have random signs. The series at fixed support length is given by $\tau_{j,\ell}=\sum_{n}\lambda^n \tau_{j,\ell}^{(n)}$ and the norms of the individual terms in this series are bounded by
\begin{equation}
    \label{Theorem7_8_1}
    \|\lambda^n \tau_{j,\ell}^{(n)}\|_2^2\geq \mathcal{B}(\ell)^{-\eta c\ell}\left(\frac{|\lambda|^2\mathcal{B}(\ell)^\eta}{C^2}\right)^n
\end{equation}
and convergence of the perturbative series thus requires that
\begin{equation}
    \label{Theorem7_9}
    |\lambda| =\left|\frac{J}{W}\right|  <C \mathcal{B}(\ell)^{-\eta/2}
\end{equation}
which is violated for every branching $\mathcal{B}(\ell)$ that diverges with support length $\ell$ and every $\eta>0$ thus proving Theorem 7. Consequently, the perturbative construction of quasilocal LIOMs fails in the thermodynamic limit in this case.

The main point of the conditional theorem 7 is that showing convergence of a scheme to perturbatively construct LIOMs requires more than just a control of resonances. It also requires control of the path entropy. Without any specific mechanism which leads to destructive interference between paths or an extreme sparsification of the operator space, the path entropy is expected to win, leading to an eventual breakdown of the perturbative scheme. Based on the exact result for the Ising chain in theorem 2, see Eq.~\eqref{bound}, which we expect to be generic for a many-body iterative commutator, the expectation is that the number of scrambling paths scales as $\mathcal{B}(\ell)\sim\ell$ (a local scrambling operator can act on any of the $\ell$ sites of the existing string). The assumption entering here is that in iterated many-body commutators almost all of the $4^\ell$ possible Pauli strings at length $\ell$ will eventually be generated unless there is a specific mechanism suppressing this growth. The symbolic calculations in Ref.~\cite{WeisseGerstnerSirker} for the disordered Heisenberg chain are consistent with such a generic exponential growth in the number of terms. It is important to stress that the obstruction considered here is not merely the static entropy of the $4^\ell$ final Pauli strings, which could in principle be compensated by a sufficiently strong exponential coefficient decay, but the path entropy that is a result of repeated fixed-support scrambling and produces a branching factor per perturbative order that itself grows with $\ell$. 

Next, we turn to the second aspect of this conditional theorem which addresses the question how different path contributions will add to each other. Without the positivity condition, which will in general not hold for the paths constituting $\tau_j^{(n)}$, the amplitudes of different scrambling paths leading to the same Pauli string are expected to add incoherently ($\eta=1$) because they are the product of random amplitudes in the microscopic Hamiltonian. Under these generic assumptions, the theorem gives the following scaling for the effective length $\ell$ where the iterative scheme starts to fail
\begin{equation}
    \label{fail}
    \ell\sim \left(\frac{W}{J}\right)^2 \, .
\end{equation}
For large disorder $W$ this is a large length scale that is out-of-reach in numerical simulations on small systems. This path-entropy argument suggests that what one sees in numerical simulations is a pre-asymptotic regime characterized by a crossover disorder strength
\begin{equation}
    \label{fail2}
    \frac{W^*(L)}{J}\sim \sqrt{L} \, .
\end{equation}
Note that this is the exact same crossover scaling that we found from a rigorous analysis of the local lower bound for the Ising chain which was a consequence of the competition between the path entropy and the exponential localization due to the growth path, see Eq.~\eqref{condition}. The exact scaling of the obstruction to the construction of LIOMs observed in numerical simulations might depend on the specific observable considered as well as on the accessible range of lengths. In particular, the linear shift of the critical disorder strength with the length of the system observed in Ref.~\cite{SuntajsBonca} for the Heisenberg chain is qualitatively not inconsistent with our findings. More generally speaking, as long as $\mathcal{B}(\ell)\to\infty$ for $\ell\to\infty$ the perturbative LIOM construction will eventually break down. 


This leaves three possibilities: (a) The conditions of theorem 7 are violated by (a1) an extreme sparsification of the operator space such that the branching $\mathcal{B}(\ell)<\mathcal{B}_0<\infty$ remains bounded, or (a2) by an almost perfect cancellation of contributions from the almost factorial number of scrambling paths, reducing their net contribution to an at most exponential growth with fixed base. At low perturbation orders, there is no indication that such a sparsification or almost perfect cancellation between different paths is taking place \cite{WeisseGerstnerSirker}. There are also no obvious algebraic structures or symmetries which could potentially be responsible for such a mechanism. If such a mechanism does exist, it would be genuinely novel and interesting physics at the heart of MBL as a stable thermodynamic phase that has not been studied so far.

(b) The quasi-local charges $\tau_j$ exist but they are not perturbatively connected to $\sigma^z_j$ for $\lambda=J/W\to 0$. A KAM multiscale renormalization scheme might then produce them for sufficiently small but finite $\lambda$ \cite{Imbrie2016,Imbrie_JSTAT,deRoeckGiacomin} but they cannot be expanded as a power series with a non-zero convergence radius in the thermodynamic limit. However, even in this case the question remains what the exact mechanism is that controls or avoids the path entropy obstruction in such a scheme. Furthermore, this type of KAM-localized phase would be quite different from the usual definition of MBL which is physically understood as originating from the strong disorder limit. 

(c) The path entropy makes the charges non-local in the thermodynamic limit and MBL as a stable thermodynamic phase simply does not exist. Here it is important to stress that this does not contradict the existence of an MBL-type regime for finite lengths $L$ because a finite length will provide a cutoff such that $\mathcal{B}(\ell)<\mathcal{B}_0(L)<\infty$ with $\ell\leq L$ and in this case a perturbative construction will converge for sufficiently small $\lambda$.

\subsection{Memory and the Mazur inequality}
Finally, we note that without any type of LIOMs and in the absence of integrability there will be no memory effect. Using Mazur's equality \cite{Mazur,Prosen,SirkerPereira,SirkerLectureNotes,PereiraPasquier} for a local operator $A_0$ and a complete set of orthogonal conserved charges $Q_j$ with $(Q_j|Q_k)=\delta_{jk}$ we find
\begin{equation}
\label{Mazur}
\lim_{T\to\infty}\frac{1}{T}\int_0^T\! \!\! \! dt\lim_{N\to\infty}\overline{\langle A_0(t)A_0\rangle}=\lim_{N\to\infty}\overline{\sum_j\frac{(A_0|Q_j)^2}{(Q_j|Q_j)}} \, ,
\end{equation}
where $\overline{\cdots}$ denotes the disorder average. If $Q_j=\tau_j$ is a set of quasi-local integrals of motion then there is one, $\tau_0$, centered near lattice site $j=0$ and $(A_0|\tau_0)\sim\mathcal{O}(1)$. In this case, memory is retained. If, however, the $Q_j$ are non-local then $(A_0|Q_j)\sim\mathcal{O}(1/\sqrt{N})$ and each individual contribution vanishes \cite{PereiraPasquier}. One might wonder if the sum over exponentially many conserved charges could still give a non-trivial result. Here, it is important to note that in the non-local case we can choose $Q_\alpha=|\alpha\rangle\langle \alpha|$ so the sum will be the diagonal ensemble. According to ETH, the diagonal ensemble converges to the thermal, infinite-temperature average which is zero for $A_0=\sigma^z_0$. I.e., the expectation for a generic disordered non-integrable chain without LIOMs is that eventually the memory of the initial state will be lost.

\section{Conclusions}
\label{Concl}
This study was motivated by the apparent tension between a rigorous result for the Ising model with random transverse and longitudinal fields showing maximal, almost factorial growth of the norm of the iterated commutator of the Hamiltonian with a local $\sigma^z$ operator \cite{Cao} and the operator-growth hypothesis stating that in an ergodic one-dimensional nearest-neighbor model the operator norm always grows at the maximal rate \cite{AvdoshkinDymarsky,ParkerCao}. This raises the question if a microscopic model can show maximal, almost factorial norm growth while remaining non-ergodic in the thermodynamic limit.

As a first step in addressing this question, we have extended the method of Ref.~\cite{Cao} to rigorously show that not only does the total operator norm $\|[H,\sigma^z]^{(k)}\|_2$ in the disordered Ising chain grow almost factorially with commutator order $k$ but that there is almost factorial weight at length $\ell_k\sim k/\ln k$. Crucially, this rigorous result goes beyond the asymptotic scaling of moments $\mu_{2k}$ and directly addresses how the operator weight evolves spatially. In particular, this local lower bound implies that the contribution of paths which remain local, i.e.~inside a fixed finite region, is asymptotically negligible; almost all of the iterated-commutator weight is generated by paths that extend to lengths of the order of the saddle point, $\ell_k\sim k/\ln k$. This rigorously excludes the strongest form of localization, dynamical locality, which is realized in the non-interacting Anderson case, where local operators remain exponentially localized. 

The mechanism for the delocalization in the many-body case is a competition between exponential localization $\e^{-\gamma\ell}$, which becomes tighter with increasing disorder strength $\gamma\sim \ln (W/J)$, and a path-entropy contribution $\ell^k$ that is independent of disorder and grows with commutator order $k$. In the asymptotic regime, the path-entropy contribution always wins. However, a finite system of length $L$ with $L\lesssim (W/J)^2$ will be in a pre-asymptotic regime where the exponential localization $\e^{-\gamma\ell}$ dominates. In the pre-asymptotic regime, the system will look Anderson-like with dynamical locality fulfilled up to path-entropy corrections which are controlled by $L$ acting as an effective infrared cutoff. This rigorous result gives a conservative scale for the onset of the asymptotic path-entropy-dominated regime and is qualitatively consistent with numerical findings that the 'critical disorder strength' shifts linearly to infinity with increasing $L$ \cite{SuntajsBonca}. At a minimum, this implies that numerical studies of small systems are ill-suited to investigate the stability of an MBL phase at strong disorder. 

Having rigorously established that almost factorial operator norm growth rules out dynamical locality, the next natural question to ask is if this maximal growth also contradicts other definitions of MBL, in particular, the MBL-LIOM definition. We have shown explicitly that the iterated commutator of a LIOM Hamiltonian with a quasi-local operator that has large off-diagonal operator content at all length scales can even show faster than $(k/\ln k)^k$ norm growth established as upper bound for microscopic nearest-neighbor models. I.e., the operator growth in a microscopic disordered spin-chain Hamiltonian and a LIOM Hamiltonian can be consistent. However, this does not imply that one can be transformed into the other; it only proves that for LIOM Hamiltonians certain quasi-local operators also show norm growth that can be consistent with the almost factorial growth in microscopic disordered spin chains. 

To address the question whether LIOMs can be constructed perturbatively for a given microscopic Hamiltonian, we have assumed that a perturbative series for the LIOMs exists, starting from the strong disorder limit, and have analytically obtained an upper bound on the convergence radius of such a series. We have shown that this convergence radius is, in general, again affected by a path-entropy obstruction. The perturbative construction of a LIOM can be viewed as a multiple commutator which, for a many-body system, is expected to contain a number of scrambling paths $\mathcal{B}(\ell)$ which diverges with the considered support length $\ell$. In particular, we expect that generically a multiple commutator will eventually generate almost all of the $4^\ell$ possible Pauli strings at length $\ell$, leading to an extensively growing number of scrambling paths $\mathcal{B}(\ell)\sim\ell$. From this argument we established a finite-size scale $L\sim (W/J)^2$ up to which a localized phase will look stable in a numerical simulation for a given disorder strength $W/J$. We note that this conditional result derived under the most natural assumptions for scrambling and the incoherent addition of amplitudes from different paths is consistent with the crossover scale that we derived rigorously based on our length-resolved result for the operator norm growth in the Ising chain. This crossover scale is also qualitatively consistent with numerical findings for the Heisenberg model \cite{SuntajsBonca, WeisseGerstnerSirker}.

We stress once more that the crossover scale $L\sim (W/J)^2$ has been rigorously established in this work for the Ising chain but is also expected to hold for other disordered spin chains such as the Heisenberg model because it relies on a path-entropy mechanism that is present in any many-body system. This implies that the question whether or not a localized phase is stable at strong disorder cannot be decisively addressed by numerical simulations. We also note that the identified path-entropy mechanism is distinct from possible instabilities based on many-body resonances and the avalanche mechanism. Overcoming the path-entropy obstruction in the perturbative construction of LIOMs would require analytical arguments for a destructive interference between $\ell^k$ scrambling paths with amplitudes that are products of random numbers or arguments for an extreme sparsification of the relevant operator space. In general, such mechanisms appear fine-tuned and are not expected in a system without any algebraic structure or symmetries enforcing it. 

A remaining open question is if iterative KAM constructions at finite $\lambda=J/W$ can avoid or overcome the path-entropy obstruction. Here, we think that it is important that the path-entropy question is explicitly addressed in any such iterative scheme and the mechanism how this obstruction is avoided is clearly identified. If the convergence of such a scheme for a finite $\lambda$ can be proven in the thermodynamic limit, it would establish a localized phase that is quite different from what is conventionally understood as MBL. In particular, such a phase would not be analytically connected to the atomic limit, thus establishing a novel phenomenology of localization. The other option is that the path-entropy obstruction is generic to any many-body system and any LIOM scheme, implying that many-body disordered spin chains are asymptotically ergodic with an upper bound on the crossover length scale between localized and non-localized that scales as $L\sim( W/J)^2$. 

\begin{acknowledgments}
The author acknowledges support  by NSERC via the Discovery grants program and gratefully acknowledges the hospitality of Utrecht University and the RPTU Kaiserslautern-Landau where part of this work was performed. The author also acknowledges helpful discussions with W.~De Roeck and F.~Huveneers.
\end{acknowledgments}


\begin{thebibliography}{49}%
\makeatletter
\providecommand \@ifxundefined [1]{%
 \@ifx{#1\undefined}
}%
\providecommand \@ifnum [1]{%
 \ifnum #1\expandafter \@firstoftwo
 \else \expandafter \@secondoftwo
 \fi
}%
\providecommand \@ifx [1]{%
 \ifx #1\expandafter \@firstoftwo
 \else \expandafter \@secondoftwo
 \fi
}%
\providecommand \natexlab [1]{#1}%
\providecommand \enquote  [1]{``#1''}%
\providecommand \bibnamefont  [1]{#1}%
\providecommand \bibfnamefont [1]{#1}%
\providecommand \citenamefont [1]{#1}%
\providecommand \href@noop [0]{\@secondoftwo}%
\providecommand \href [0]{\begingroup \@sanitize@url \@href}%
\providecommand \@href[1]{\@@startlink{#1}\@@href}%
\providecommand \@@href[1]{\endgroup#1\@@endlink}%
\providecommand \@sanitize@url [0]{\catcode `\\12\catcode `\$12\catcode
  `\&12\catcode `\#12\catcode `\^12\catcode `\_12\catcode `\%12\relax}%
\providecommand \@@startlink[1]{}%
\providecommand \@@endlink[0]{}%
\providecommand \url  [0]{\begingroup\@sanitize@url \@url }%
\providecommand \@url [1]{\endgroup\@href {#1}{\urlprefix }}%
\providecommand \urlprefix  [0]{URL }%
\providecommand \Eprint [0]{\href }%
\providecommand \doibase [0]{https://doi.org/}%
\providecommand \selectlanguage [0]{\@gobble}%
\providecommand \bibinfo  [0]{\@secondoftwo}%
\providecommand \bibfield  [0]{\@secondoftwo}%
\providecommand \translation [1]{[#1]}%
\providecommand \BibitemOpen [0]{}%
\providecommand \bibitemStop [0]{}%
\providecommand \bibitemNoStop [0]{.\EOS\space}%
\providecommand \EOS [0]{\spacefactor3000\relax}%
\providecommand \BibitemShut  [1]{\csname bibitem#1\endcsname}%
\let\auto@bib@innerbib\@empty
\bibitem [{\citenamefont {Cao}(2021)}]{Cao}%
  \BibitemOpen
  \bibfield  {author} {\bibinfo {author} {\bibfnamefont {X.}~\bibnamefont
  {Cao}},\ }\bibfield  {title} {\bibinfo {title} {A statistical mechanism for
  operator growth},\ }\href {https://doi.org/10.1088/1751-8121/abe77c}
  {\bibfield  {journal} {\bibinfo  {journal} {J. Phys. A}\ }\textbf {\bibinfo
  {volume} {54}},\ \bibinfo {pages} {144001} (\bibinfo {year}
  {2021})}\BibitemShut {NoStop}%
\bibitem [{\citenamefont {Anderson}(1958)}]{Anderson58}%
  \BibitemOpen
  \bibfield  {author} {\bibinfo {author} {\bibfnamefont {P.~W.}\ \bibnamefont
  {Anderson}},\ }\bibfield  {title} {\bibinfo {title} {Absence of diffusion in
  certain random lattices},\ }\href {https://doi.org/10.1103/PhysRev.109.1492}
  {\bibfield  {journal} {\bibinfo  {journal} {Phys. Rev.}\ }\textbf {\bibinfo
  {volume} {109}},\ \bibinfo {pages} {1492} (\bibinfo {year}
  {1958})}\BibitemShut {NoStop}%
\bibitem [{\citenamefont {Abrahams}\ \emph {et~al.}(1979)\citenamefont
  {Abrahams}, \citenamefont {Anderson}, \citenamefont {Licciardello},\ and\
  \citenamefont {Ramakrishnan}}]{AbrahamsAnderson}%
  \BibitemOpen
  \bibfield  {author} {\bibinfo {author} {\bibfnamefont {E.}~\bibnamefont
  {Abrahams}}, \bibinfo {author} {\bibfnamefont {P.~W.}\ \bibnamefont
  {Anderson}}, \bibinfo {author} {\bibfnamefont {D.~C.}\ \bibnamefont
  {Licciardello}},\ and\ \bibinfo {author} {\bibfnamefont {T.~V.}\ \bibnamefont
  {Ramakrishnan}},\ }\bibfield  {title} {\bibinfo {title} {Scaling theory of
  localization: Absence of quantum diffusion in two dimensions},\ }\href
  {https://doi.org/10.1103/PhysRevLett.42.673} {\bibfield  {journal} {\bibinfo
  {journal} {Phys. Rev. Lett.}\ }\textbf {\bibinfo {volume} {42}},\ \bibinfo
  {pages} {673} (\bibinfo {year} {1979})}\BibitemShut {NoStop}%
\bibitem [{\citenamefont {Edwards}\ and\ \citenamefont
  {Thouless}(1972)}]{EdwardsThouless}%
  \BibitemOpen
  \bibfield  {author} {\bibinfo {author} {\bibfnamefont {J.~T.}\ \bibnamefont
  {Edwards}}\ and\ \bibinfo {author} {\bibfnamefont {D.~J.}\ \bibnamefont
  {Thouless}},\ }\bibfield  {title} {\bibinfo {title} {Numerical studies of
  localization in disordered systems},\ }\href
  {http://stacks.iop.org/0022-3719/5/i=8/a=007} {\bibfield  {journal} {\bibinfo
   {journal} {J. Phys. C}\ }\textbf {\bibinfo {volume} {5}},\ \bibinfo {pages}
  {807} (\bibinfo {year} {1972})}\BibitemShut {NoStop}%
\bibitem [{\citenamefont {Bucaj}\ \emph {et~al.}(2019)\citenamefont {Bucaj},
  \citenamefont {Damanik}, \citenamefont {Fillman}, \citenamefont {Gerbuz},
  \citenamefont {VandenBoom}, \citenamefont {Wang},\ and\ \citenamefont
  {Zhang}}]{BucajDamanik}%
  \BibitemOpen
  \bibfield  {author} {\bibinfo {author} {\bibfnamefont {V.}~\bibnamefont
  {Bucaj}}, \bibinfo {author} {\bibfnamefont {D.}~\bibnamefont {Damanik}},
  \bibinfo {author} {\bibfnamefont {J.}~\bibnamefont {Fillman}}, \bibinfo
  {author} {\bibfnamefont {V.}~\bibnamefont {Gerbuz}}, \bibinfo {author}
  {\bibfnamefont {T.}~\bibnamefont {VandenBoom}}, \bibinfo {author}
  {\bibfnamefont {F.}~\bibnamefont {Wang}},\ and\ \bibinfo {author}
  {\bibfnamefont {Z.}~\bibnamefont {Zhang}},\ }\bibfield  {title} {\bibinfo
  {title} {Localization for the one-dimensional {Anderson} model via positivity
  and large deviations for the {Lyapunov} exponent},\ }\href
  {https://doi.org/10.1090/tran/7832} {\bibfield  {journal} {\bibinfo
  {journal} {Transactions of the American Mathematical Society}\ }\textbf
  {\bibinfo {volume} {372}},\ \bibinfo {pages} {3619} (\bibinfo {year}
  {2019})}\BibitemShut {NoStop}%
\bibitem [{\citenamefont {Hu}\ \emph {et~al.}(2008)\citenamefont {Hu},
  \citenamefont {Strybulevych}, \citenamefont {Page}, \citenamefont
  {Skipetrov},\ and\ \citenamefont {van Tiggelen}}]{Page2008}%
  \BibitemOpen
  \bibfield  {author} {\bibinfo {author} {\bibfnamefont {H.}~\bibnamefont
  {Hu}}, \bibinfo {author} {\bibfnamefont {A.}~\bibnamefont {Strybulevych}},
  \bibinfo {author} {\bibfnamefont {J.~H.}\ \bibnamefont {Page}}, \bibinfo
  {author} {\bibfnamefont {S.~E.}\ \bibnamefont {Skipetrov}},\ and\ \bibinfo
  {author} {\bibfnamefont {B.~A.}\ \bibnamefont {van Tiggelen}},\ }\bibfield
  {title} {\bibinfo {title} {Localization of ultrasound in a three-dimensional
  elastic network},\ }\href {https://doi.org/10.1038/nphys1101} {\bibfield
  {journal} {\bibinfo  {journal} {Nature Physics}\ }\textbf {\bibinfo {volume}
  {4}},\ \bibinfo {pages} {945} (\bibinfo {year} {2008})}\BibitemShut {NoStop}%
\bibitem [{\citenamefont {Basko}\ \emph {et~al.}(2006)\citenamefont {Basko},
  \citenamefont {Aleiner},\ and\ \citenamefont {Altshuler}}]{BaskoAleiner}%
  \BibitemOpen
  \bibfield  {author} {\bibinfo {author} {\bibfnamefont {D.~M.}\ \bibnamefont
  {Basko}}, \bibinfo {author} {\bibfnamefont {I.~L.}\ \bibnamefont {Aleiner}},\
  and\ \bibinfo {author} {\bibfnamefont {B.~L.}\ \bibnamefont {Altshuler}},\
  }\bibfield  {title} {\bibinfo {title} {Metal-insulator transition in a weakly
  interacting many-electron system with localized single-particle states},\
  }\href {https://doi.org/https://doi.org/10.1016/j.aop.2005.11.014} {\bibfield
   {journal} {\bibinfo  {journal} {Ann. Phys.}\ }\textbf {\bibinfo {volume}
  {321}},\ \bibinfo {pages} {1126} (\bibinfo {year} {2006})}\BibitemShut
  {NoStop}%
\bibitem [{\citenamefont {Oganesyan}\ and\ \citenamefont
  {Huse}(2007)}]{OganesyanHuse}%
  \BibitemOpen
  \bibfield  {author} {\bibinfo {author} {\bibfnamefont {V.}~\bibnamefont
  {Oganesyan}}\ and\ \bibinfo {author} {\bibfnamefont {D.~A.}\ \bibnamefont
  {Huse}},\ }\bibfield  {title} {\bibinfo {title} {Localization of interacting
  fermions at high temperature},\ }\href
  {https://doi.org/10.1103/PhysRevB.75.155111} {\bibfield  {journal} {\bibinfo
  {journal} {Phys. Rev. B}\ }\textbf {\bibinfo {volume} {75}},\ \bibinfo
  {pages} {155111} (\bibinfo {year} {2007})}\BibitemShut {NoStop}%
\bibitem [{\citenamefont {Pal}\ and\ \citenamefont {Huse}(2010)}]{PalHuse}%
  \BibitemOpen
  \bibfield  {author} {\bibinfo {author} {\bibfnamefont {A.}~\bibnamefont
  {Pal}}\ and\ \bibinfo {author} {\bibfnamefont {D.~A.}\ \bibnamefont {Huse}},\
  }\bibfield  {title} {\bibinfo {title} {Many-body localization phase
  transition},\ }\href {https://doi.org/10.1103/PhysRevB.82.174411} {\bibfield
  {journal} {\bibinfo  {journal} {Phys. Rev. B}\ }\textbf {\bibinfo {volume}
  {82}},\ \bibinfo {pages} {174411} (\bibinfo {year} {2010})}\BibitemShut
  {NoStop}%
\bibitem [{\citenamefont {\ifmmode \check{Z}\else
  \v{Z}\fi{}nidari\ifmmode~\check{c}\else \v{c}\fi{}}\ \emph
  {et~al.}(2008)\citenamefont {\ifmmode \check{Z}\else
  \v{Z}\fi{}nidari\ifmmode~\check{c}\else \v{c}\fi{}}, \citenamefont {Prosen},\
  and\ \citenamefont {Prelov\ifmmode~\check{s}\else
  \v{s}\fi{}ek}}]{ZnidaricProsen}%
  \BibitemOpen
  \bibfield  {author} {\bibinfo {author} {\bibfnamefont {M.}~\bibnamefont
  {\ifmmode \check{Z}\else \v{Z}\fi{}nidari\ifmmode~\check{c}\else
  \v{c}\fi{}}}, \bibinfo {author} {\bibfnamefont {T.}~\bibnamefont {Prosen}},\
  and\ \bibinfo {author} {\bibfnamefont {P.}~\bibnamefont
  {Prelov\ifmmode~\check{s}\else \v{s}\fi{}ek}},\ }\bibfield  {title} {\bibinfo
  {title} {Many-body localization in the heisenberg $xxz$ magnet in a random
  field},\ }\href {https://doi.org/10.1103/PhysRevB.77.064426} {\bibfield
  {journal} {\bibinfo  {journal} {Phys. Rev. B}\ }\textbf {\bibinfo {volume}
  {77}},\ \bibinfo {pages} {064426} (\bibinfo {year} {2008})}\BibitemShut
  {NoStop}%
\bibitem [{\citenamefont {Bardarson}\ \emph {et~al.}(2012)\citenamefont
  {Bardarson}, \citenamefont {Pollmann},\ and\ \citenamefont
  {Moore}}]{BardarsonPollmann}%
  \BibitemOpen
  \bibfield  {author} {\bibinfo {author} {\bibfnamefont {J.~H.}\ \bibnamefont
  {Bardarson}}, \bibinfo {author} {\bibfnamefont {F.}~\bibnamefont
  {Pollmann}},\ and\ \bibinfo {author} {\bibfnamefont {J.~E.}\ \bibnamefont
  {Moore}},\ }\bibfield  {title} {\bibinfo {title} {Unbounded growth of
  entanglement in models of many-body localization},\ }\href
  {https://doi.org/10.1103/PhysRevLett.109.017202} {\bibfield  {journal}
  {\bibinfo  {journal} {Phys. Rev. Lett.}\ }\textbf {\bibinfo {volume} {109}},\
  \bibinfo {pages} {017202} (\bibinfo {year} {2012})}\BibitemShut {NoStop}%
\bibitem [{\citenamefont {Schreiber}\ \emph {et~al.}(2015)\citenamefont
  {Schreiber}, \citenamefont {Hodgman}, \citenamefont {Bordia}, \citenamefont
  {L{\"u}schen}, \citenamefont {Fischer}, \citenamefont {Vosk}, \citenamefont
  {Altman}, \citenamefont {Schneider},\ and\ \citenamefont
  {Bloch}}]{SchreiberHodgman}%
  \BibitemOpen
  \bibfield  {author} {\bibinfo {author} {\bibfnamefont {M.}~\bibnamefont
  {Schreiber}}, \bibinfo {author} {\bibfnamefont {S.~S.}\ \bibnamefont
  {Hodgman}}, \bibinfo {author} {\bibfnamefont {P.}~\bibnamefont {Bordia}},
  \bibinfo {author} {\bibfnamefont {H.~P.}\ \bibnamefont {L{\"u}schen}},
  \bibinfo {author} {\bibfnamefont {M.~H.}\ \bibnamefont {Fischer}}, \bibinfo
  {author} {\bibfnamefont {R.}~\bibnamefont {Vosk}}, \bibinfo {author}
  {\bibfnamefont {E.}~\bibnamefont {Altman}}, \bibinfo {author} {\bibfnamefont
  {U.}~\bibnamefont {Schneider}},\ and\ \bibinfo {author} {\bibfnamefont
  {I.}~\bibnamefont {Bloch}},\ }\bibfield  {title} {\bibinfo {title}
  {Observation of many-body localization of interacting fermions in a
  quasi-random optical lattice},\ }\href
  {https://doi.org/10.1126/science.aaa7432} {\bibfield  {journal} {\bibinfo
  {journal} {Science}\ }\textbf {\bibinfo {volume} {349}},\ \bibinfo {pages}
  {842} (\bibinfo {year} {2015})}\BibitemShut {NoStop}%
\bibitem [{\citenamefont {Nandkishore}\ and\ \citenamefont
  {Huse}(2015)}]{NandkishoreHuse}%
  \BibitemOpen
  \bibfield  {author} {\bibinfo {author} {\bibfnamefont {R.}~\bibnamefont
  {Nandkishore}}\ and\ \bibinfo {author} {\bibfnamefont {D.}~\bibnamefont
  {Huse}},\ }\bibfield  {title} {\bibinfo {title} {Many-body localization and
  thermalization in quantum statistical mechanics},\ }\href
  {https://doi.org/10.1146/annurev-conmatphys-031214-014726} {\bibfield
  {journal} {\bibinfo  {journal} {Ann. Rev. Cond. Mat. Phys.}\ }\textbf
  {\bibinfo {volume} {6}},\ \bibinfo {pages} {15} (\bibinfo {year}
  {2015})}\BibitemShut {NoStop}%
\bibitem [{\citenamefont {Altman}\ and\ \citenamefont
  {Vosk}(2015)}]{AltmanVoskReview}%
  \BibitemOpen
  \bibfield  {author} {\bibinfo {author} {\bibfnamefont {E.}~\bibnamefont
  {Altman}}\ and\ \bibinfo {author} {\bibfnamefont {R.}~\bibnamefont {Vosk}},\
  }\bibfield  {title} {\bibinfo {title} {Universal dynamics and renormalization
  in many-body-localized systems},\ }\href
  {https://doi.org/10.1146/annurev-conmatphys-031214-014701} {\bibfield
  {journal} {\bibinfo  {journal} {Annual Review of Condensed Matter Physics}\
  }\textbf {\bibinfo {volume} {6}},\ \bibinfo {pages} {383} (\bibinfo {year}
  {2015})}\BibitemShut {NoStop}%
\bibitem [{\citenamefont {Weisse}\ \emph {et~al.}(2025)\citenamefont {Weisse},
  \citenamefont {Gerstner},\ and\ \citenamefont
  {Sirker}}]{WeisseGerstnerSirker}%
  \BibitemOpen
  \bibfield  {author} {\bibinfo {author} {\bibfnamefont {A.}~\bibnamefont
  {Weisse}}, \bibinfo {author} {\bibfnamefont {R.}~\bibnamefont {Gerstner}},\
  and\ \bibinfo {author} {\bibfnamefont {J.}~\bibnamefont {Sirker}},\
  }\bibfield  {title} {\bibinfo {title} {Operator growth in disordered spin
  chains: Indications for the absence of many-body localization},\ }\href
  {https://doi.org/10.1103/wgss-nt8t} {\bibfield  {journal} {\bibinfo
  {journal} {Phys. Rev. Res.}\ }\textbf {\bibinfo {volume} {7}},\ \bibinfo
  {pages} {033018} (\bibinfo {year} {2025})}\BibitemShut {NoStop}%
\bibitem [{\citenamefont {Sels}\ and\ \citenamefont
  {Polkovnikov}(2021)}]{SelsPolkovnikov}%
  \BibitemOpen
  \bibfield  {author} {\bibinfo {author} {\bibfnamefont {D.}~\bibnamefont
  {Sels}}\ and\ \bibinfo {author} {\bibfnamefont {A.}~\bibnamefont
  {Polkovnikov}},\ }\bibfield  {title} {\bibinfo {title} {Dynamical obstruction
  to localization in a disordered spin chain},\ }\href
  {https://doi.org/10.1103/PhysRevE.104.054105} {\bibfield  {journal} {\bibinfo
   {journal} {Phys. Rev. E}\ }\textbf {\bibinfo {volume} {104}},\ \bibinfo
  {pages} {054105} (\bibinfo {year} {2021})}\BibitemShut {NoStop}%
\bibitem [{\citenamefont {Sels}\ and\ \citenamefont
  {Polkovnikov}(2023)}]{SelsPolkovnikovPRX}%
  \BibitemOpen
  \bibfield  {author} {\bibinfo {author} {\bibfnamefont {D.}~\bibnamefont
  {Sels}}\ and\ \bibinfo {author} {\bibfnamefont {A.}~\bibnamefont
  {Polkovnikov}},\ }\bibfield  {title} {\bibinfo {title} {Thermalization of
  dilute impurities in one-dimensional spin chains},\ }\href
  {https://doi.org/10.1103/PhysRevX.13.011041} {\bibfield  {journal} {\bibinfo
  {journal} {Phys. Rev. X}\ }\textbf {\bibinfo {volume} {13}},\ \bibinfo
  {pages} {011041} (\bibinfo {year} {2023})}\BibitemShut {NoStop}%
\bibitem [{\citenamefont {Abanin}\ \emph {et~al.}(2019)\citenamefont {Abanin},
  \citenamefont {Altman}, \citenamefont {Bloch},\ and\ \citenamefont
  {Serbyn}}]{AbaninRev2019}%
  \BibitemOpen
  \bibfield  {author} {\bibinfo {author} {\bibfnamefont {D.~A.}\ \bibnamefont
  {Abanin}}, \bibinfo {author} {\bibfnamefont {E.}~\bibnamefont {Altman}},
  \bibinfo {author} {\bibfnamefont {I.}~\bibnamefont {Bloch}},\ and\ \bibinfo
  {author} {\bibfnamefont {M.}~\bibnamefont {Serbyn}},\ }\bibfield  {title}
  {\bibinfo {title} {Colloquium: Many-body localization, thermalization, and
  entanglement},\ }\href {https://doi.org/10.1103/RevModPhys.91.021001}
  {\bibfield  {journal} {\bibinfo  {journal} {Rev. Mod. Phys.}\ }\textbf
  {\bibinfo {volume} {91}},\ \bibinfo {pages} {021001} (\bibinfo {year}
  {2019})}\BibitemShut {NoStop}%
\bibitem [{\citenamefont {Kiefer-Emmanouilidis}\ \emph
  {et~al.}(2020)\citenamefont {Kiefer-Emmanouilidis}, \citenamefont {Unanyan},
  \citenamefont {Fleischhauer},\ and\ \citenamefont {Sirker}}]{KieferUnanyan2}%
  \BibitemOpen
  \bibfield  {author} {\bibinfo {author} {\bibfnamefont {M.}~\bibnamefont
  {Kiefer-Emmanouilidis}}, \bibinfo {author} {\bibfnamefont {R.}~\bibnamefont
  {Unanyan}}, \bibinfo {author} {\bibfnamefont {M.}~\bibnamefont
  {Fleischhauer}},\ and\ \bibinfo {author} {\bibfnamefont {J.}~\bibnamefont
  {Sirker}},\ }\bibfield  {title} {\bibinfo {title} {Evidence for unbounded
  growth of the number entropy in many-body localized phases},\ }\href@noop {}
  {\bibfield  {journal} {\bibinfo  {journal} {Phys. Rev. Lett.}\ }\textbf
  {\bibinfo {volume} {124}},\ \bibinfo {pages} {243601} (\bibinfo {year}
  {2020})}\BibitemShut {NoStop}%
\bibitem [{\citenamefont {Kiefer-Emmanouilidis}\ \emph
  {et~al.}(2021)\citenamefont {Kiefer-Emmanouilidis}, \citenamefont {Unanyan},
  \citenamefont {Fleischhauer},\ and\ \citenamefont {Sirker}}]{KieferUnanyan3}%
  \BibitemOpen
  \bibfield  {author} {\bibinfo {author} {\bibfnamefont {M.}~\bibnamefont
  {Kiefer-Emmanouilidis}}, \bibinfo {author} {\bibfnamefont {R.}~\bibnamefont
  {Unanyan}}, \bibinfo {author} {\bibfnamefont {M.}~\bibnamefont
  {Fleischhauer}},\ and\ \bibinfo {author} {\bibfnamefont {J.}~\bibnamefont
  {Sirker}},\ }\bibfield  {title} {\bibinfo {title} {Slow delocalization of
  particles in many-body localized phases},\ }\href
  {https://doi.org/10.1103/PhysRevB.103.024203} {\bibfield  {journal} {\bibinfo
   {journal} {Phys. Rev. B}\ }\textbf {\bibinfo {volume} {103}},\ \bibinfo
  {pages} {024203} (\bibinfo {year} {2021})}\BibitemShut {NoStop}%
\bibitem [{\citenamefont {Suntajs}\ \emph {et~al.}(2020)\citenamefont
  {Suntajs}, \citenamefont {Bonca}, \citenamefont {Prosen},\ and\ \citenamefont
  {Vidmar}}]{SuntajsBonca}%
  \BibitemOpen
  \bibfield  {author} {\bibinfo {author} {\bibfnamefont {J.}~\bibnamefont
  {Suntajs}}, \bibinfo {author} {\bibfnamefont {J.}~\bibnamefont {Bonca}},
  \bibinfo {author} {\bibfnamefont {T.}~\bibnamefont {Prosen}},\ and\ \bibinfo
  {author} {\bibfnamefont {L.}~\bibnamefont {Vidmar}},\ }\bibfield  {title}
  {\bibinfo {title} {Quantum chaos challenges many-body localization},\ }\href
  {https://doi.org/10.1103/PhysRevE.102.062144} {\bibfield  {journal} {\bibinfo
   {journal} {Phys. Rev. E}\ }\textbf {\bibinfo {volume} {102}},\ \bibinfo
  {pages} {062144} (\bibinfo {year} {2020})}\BibitemShut {NoStop}%
\bibitem [{\citenamefont {De~Roeck}\ \emph {et~al.}(2023)\citenamefont
  {De~Roeck}, \citenamefont {Huveneers}, \citenamefont {Meeus},\ and\
  \citenamefont {Pro{\'s}niak}}]{DeRoeckQuasilocalization}%
  \BibitemOpen
  \bibfield  {author} {\bibinfo {author} {\bibfnamefont {W.}~\bibnamefont
  {De~Roeck}}, \bibinfo {author} {\bibfnamefont {F.}~\bibnamefont {Huveneers}},
  \bibinfo {author} {\bibfnamefont {B.}~\bibnamefont {Meeus}},\ and\ \bibinfo
  {author} {\bibfnamefont {O.~A.}\ \bibnamefont {Pro{\'s}niak}},\ }\bibfield
  {title} {\bibinfo {title} {Rigorous and simple results on very slow
  thermalization, or quasi-localization, of the disordered quantum chain},\
  }\href {https://doi.org/10.1016/j.physa.2023.129245} {\bibfield  {journal}
  {\bibinfo  {journal} {Physica A: Statistical Mechanics and its Applications}\
  }\textbf {\bibinfo {volume} {631}},\ \bibinfo {pages} {129245} (\bibinfo
  {year} {2023})}\BibitemShut {NoStop}%
\bibitem [{\citenamefont {Luitz}\ \emph {et~al.}(2015)\citenamefont {Luitz},
  \citenamefont {Laflorencie},\ and\ \citenamefont {Alet}}]{Luitz1}%
  \BibitemOpen
  \bibfield  {author} {\bibinfo {author} {\bibfnamefont {D.~J.}\ \bibnamefont
  {Luitz}}, \bibinfo {author} {\bibfnamefont {N.}~\bibnamefont {Laflorencie}},\
  and\ \bibinfo {author} {\bibfnamefont {F.}~\bibnamefont {Alet}},\ }\bibfield
  {title} {\bibinfo {title} {Many-body localization edge in the random-field
  heisenberg chain},\ }\href {https://doi.org/10.1103/PhysRevB.91.081103}
  {\bibfield  {journal} {\bibinfo  {journal} {Phys. Rev. B}\ }\textbf {\bibinfo
  {volume} {91}},\ \bibinfo {pages} {081103} (\bibinfo {year}
  {2015})}\BibitemShut {NoStop}%
\bibitem [{\citenamefont {Luitz}\ \emph {et~al.}(2016)\citenamefont {Luitz},
  \citenamefont {Laflorencie},\ and\ \citenamefont {Alet}}]{Luitz2}%
  \BibitemOpen
  \bibfield  {author} {\bibinfo {author} {\bibfnamefont {D.~J.}\ \bibnamefont
  {Luitz}}, \bibinfo {author} {\bibfnamefont {N.}~\bibnamefont {Laflorencie}},\
  and\ \bibinfo {author} {\bibfnamefont {F.}~\bibnamefont {Alet}},\ }\bibfield
  {title} {\bibinfo {title} {Extended slow dynamical regime close to the
  many-body localization transition},\ }\href
  {https://doi.org/10.1103/PhysRevB.93.060201} {\bibfield  {journal} {\bibinfo
  {journal} {Phys. Rev. B}\ }\textbf {\bibinfo {volume} {93}},\ \bibinfo
  {pages} {060201} (\bibinfo {year} {2016})}\BibitemShut {NoStop}%
\bibitem [{\citenamefont {Sierant}\ \emph {et~al.}(2020)\citenamefont
  {Sierant}, \citenamefont {Delande},\ and\ \citenamefont
  {Zakrzewski}}]{SierantDelande}%
  \BibitemOpen
  \bibfield  {author} {\bibinfo {author} {\bibfnamefont {P.}~\bibnamefont
  {Sierant}}, \bibinfo {author} {\bibfnamefont {D.}~\bibnamefont {Delande}},\
  and\ \bibinfo {author} {\bibfnamefont {J.}~\bibnamefont {Zakrzewski}},\
  }\bibfield  {title} {\bibinfo {title} {Thouless time analysis of anderson and
  many-body localization transitions},\ }\href
  {https://doi.org/10.1103/PhysRevLett.124.186601} {\bibfield  {journal}
  {\bibinfo  {journal} {Phys. Rev. Lett.}\ }\textbf {\bibinfo {volume} {124}},\
  \bibinfo {pages} {186601} (\bibinfo {year} {2020})}\BibitemShut {NoStop}%
\bibitem [{\citenamefont {Abanin}\ \emph {et~al.}(2021)\citenamefont {Abanin},
  \citenamefont {Bardarson}, \citenamefont {{De Tomasi}}, \citenamefont
  {Gopalakrishnan}, \citenamefont {Khemani}, \citenamefont {Parameswaran},
  \citenamefont {Pollmann}, \citenamefont {Potter}, \citenamefont {Serbyn},\
  and\ \citenamefont {Vasseur}}]{AbaninBardarson}%
  \BibitemOpen
  \bibfield  {author} {\bibinfo {author} {\bibfnamefont {D.}~\bibnamefont
  {Abanin}}, \bibinfo {author} {\bibfnamefont {J.}~\bibnamefont {Bardarson}},
  \bibinfo {author} {\bibfnamefont {G.}~\bibnamefont {{De Tomasi}}}, \bibinfo
  {author} {\bibfnamefont {S.}~\bibnamefont {Gopalakrishnan}}, \bibinfo
  {author} {\bibfnamefont {V.}~\bibnamefont {Khemani}}, \bibinfo {author}
  {\bibfnamefont {S.}~\bibnamefont {Parameswaran}}, \bibinfo {author}
  {\bibfnamefont {F.}~\bibnamefont {Pollmann}}, \bibinfo {author}
  {\bibfnamefont {A.}~\bibnamefont {Potter}}, \bibinfo {author} {\bibfnamefont
  {M.}~\bibnamefont {Serbyn}},\ and\ \bibinfo {author} {\bibfnamefont
  {R.}~\bibnamefont {Vasseur}},\ }\bibfield  {title} {\bibinfo {title}
  {Distinguishing localization from chaos: Challenges in finite-size systems},\
  }\href {https://doi.org/https://doi.org/10.1016/j.aop.2021.168415} {\bibfield
   {journal} {\bibinfo  {journal} {Annals of Physics}\ }\textbf {\bibinfo
  {volume} {427}},\ \bibinfo {pages} {168415} (\bibinfo {year}
  {2021})}\BibitemShut {NoStop}%
\bibitem [{\citenamefont {Serbyn}\ \emph {et~al.}(2013)\citenamefont {Serbyn},
  \citenamefont {Papi\ifmmode~\acute{c}\else \'{c}\fi{}},\ and\ \citenamefont
  {Abanin}}]{SerbynPapic}%
  \BibitemOpen
  \bibfield  {author} {\bibinfo {author} {\bibfnamefont {M.}~\bibnamefont
  {Serbyn}}, \bibinfo {author} {\bibfnamefont {Z.}~\bibnamefont
  {Papi\ifmmode~\acute{c}\else \'{c}\fi{}}},\ and\ \bibinfo {author}
  {\bibfnamefont {D.~A.}\ \bibnamefont {Abanin}},\ }\bibfield  {title}
  {\bibinfo {title} {Local conservation laws and the structure of the many-body
  localized states},\ }\href {https://doi.org/10.1103/PhysRevLett.111.127201}
  {\bibfield  {journal} {\bibinfo  {journal} {Phys. Rev. Lett.}\ }\textbf
  {\bibinfo {volume} {111}},\ \bibinfo {pages} {127201} (\bibinfo {year}
  {2013})}\BibitemShut {NoStop}%
\bibitem [{\citenamefont {Ros}\ \emph {et~al.}(2014)\citenamefont {Ros},
  \citenamefont {M\"uller},\ and\ \citenamefont
  {Scardicchio}}]{RosMuellerScardicchio}%
  \BibitemOpen
  \bibfield  {author} {\bibinfo {author} {\bibfnamefont {V.}~\bibnamefont
  {Ros}}, \bibinfo {author} {\bibfnamefont {M.}~\bibnamefont {M\"uller}},\ and\
  \bibinfo {author} {\bibfnamefont {A.}~\bibnamefont {Scardicchio}},\
  }\bibfield  {title} {\bibinfo {title} {Integrals of motion in the many-body
  localized phase},\ }\href
  {https://doi.org/https://doi.org/10.1016/j.nuclphysb.2014.12.014} {\bibfield
  {journal} {\bibinfo  {journal} {Nucl. Phys. B}\ }\textbf {\bibinfo {volume}
  {891}},\ \bibinfo {pages} {420} (\bibinfo {year} {2014})}\BibitemShut
  {NoStop}%
\bibitem [{\citenamefont {Chandran}\ \emph {et~al.}(2015)\citenamefont
  {Chandran}, \citenamefont {Kim}, \citenamefont {Vidal},\ and\ \citenamefont
  {Abanin}}]{ChandranKim}%
  \BibitemOpen
  \bibfield  {author} {\bibinfo {author} {\bibfnamefont {A.}~\bibnamefont
  {Chandran}}, \bibinfo {author} {\bibfnamefont {I.~H.}\ \bibnamefont {Kim}},
  \bibinfo {author} {\bibfnamefont {G.}~\bibnamefont {Vidal}},\ and\ \bibinfo
  {author} {\bibfnamefont {D.~A.}\ \bibnamefont {Abanin}},\ }\bibfield  {title}
  {\bibinfo {title} {Constructing local integrals of motion in the many-body
  localized phase},\ }\href {https://doi.org/10.1103/PhysRevB.91.085425}
  {\bibfield  {journal} {\bibinfo  {journal} {Phys. Rev. B}\ }\textbf {\bibinfo
  {volume} {91}},\ \bibinfo {pages} {085425} (\bibinfo {year}
  {2015})}\BibitemShut {NoStop}%
\bibitem [{\citenamefont {Huse}\ \emph {et~al.}(2014)\citenamefont {Huse},
  \citenamefont {Nandkishore},\ and\ \citenamefont
  {Oganesyan}}]{HuseNandkishore}%
  \BibitemOpen
  \bibfield  {author} {\bibinfo {author} {\bibfnamefont {D.~A.}\ \bibnamefont
  {Huse}}, \bibinfo {author} {\bibfnamefont {R.}~\bibnamefont {Nandkishore}},\
  and\ \bibinfo {author} {\bibfnamefont {V.}~\bibnamefont {Oganesyan}},\
  }\bibfield  {title} {\bibinfo {title} {Phenomenology of fully
  many-body-localized systems},\ }\href
  {https://doi.org/10.1103/PhysRevB.90.174202} {\bibfield  {journal} {\bibinfo
  {journal} {Phys. Rev. B}\ }\textbf {\bibinfo {volume} {90}},\ \bibinfo
  {pages} {174202} (\bibinfo {year} {2014})}\BibitemShut {NoStop}%
\bibitem [{\citenamefont {Imbrie}\ \emph {et~al.}(2017)\citenamefont {Imbrie},
  \citenamefont {Ros},\ and\ \citenamefont
  {Scardicchio}}]{ImbrieRosScardicchio}%
  \BibitemOpen
  \bibfield  {author} {\bibinfo {author} {\bibfnamefont {J.~Z.}\ \bibnamefont
  {Imbrie}}, \bibinfo {author} {\bibfnamefont {V.}~\bibnamefont {Ros}},\ and\
  \bibinfo {author} {\bibfnamefont {A.}~\bibnamefont {Scardicchio}},\
  }\bibfield  {title} {\bibinfo {title} {Local integrals of motion in many-body
  localized systems},\ }\href {https://doi.org/10.1002/andp.201600278}
  {\bibfield  {journal} {\bibinfo  {journal} {Annalen der Physik}\ }\textbf
  {\bibinfo {volume} {529}},\ \bibinfo {pages} {1600278} (\bibinfo {year}
  {2017})}\BibitemShut {NoStop}%
\bibitem [{\citenamefont {Imbrie}(2016{\natexlab{a}})}]{Imbrie_JSTAT}%
  \BibitemOpen
  \bibfield  {author} {\bibinfo {author} {\bibfnamefont {J.}~\bibnamefont
  {Imbrie}},\ }\bibfield  {title} {\bibinfo {title} {On many-body localization
  for quantum spin chains},\ }\href {https://doi.org/10.1007/s10955-016-1508-x}
  {\bibfield  {journal} {\bibinfo  {journal} {J. Stat. Phys.}\ }\textbf
  {\bibinfo {volume} {163}},\ \bibinfo {pages} {998} (\bibinfo {year}
  {2016}{\natexlab{a}})}\BibitemShut {NoStop}%
\bibitem [{\citenamefont {Avdoshkin}\ and\ \citenamefont
  {Dymarsky}(2020)}]{AvdoshkinDymarsky}%
  \BibitemOpen
  \bibfield  {author} {\bibinfo {author} {\bibfnamefont {A.}~\bibnamefont
  {Avdoshkin}}\ and\ \bibinfo {author} {\bibfnamefont {A.}~\bibnamefont
  {Dymarsky}},\ }\bibfield  {title} {\bibinfo {title} {Euclidean operator
  growth and quantum chaos},\ }\href
  {https://doi.org/10.1103/PhysRevResearch.2.043234} {\bibfield  {journal}
  {\bibinfo  {journal} {Phys. Rev. Res.}\ }\textbf {\bibinfo {volume} {2}},\
  \bibinfo {pages} {043234} (\bibinfo {year} {2020})}\BibitemShut {NoStop}%
\bibitem [{\citenamefont {Parker}\ \emph {et~al.}(2019)\citenamefont {Parker},
  \citenamefont {Cao}, \citenamefont {Avdoshkin}, \citenamefont {Scaffidi},\
  and\ \citenamefont {Altman}}]{ParkerCao}%
  \BibitemOpen
  \bibfield  {author} {\bibinfo {author} {\bibfnamefont {D.~E.}\ \bibnamefont
  {Parker}}, \bibinfo {author} {\bibfnamefont {X.}~\bibnamefont {Cao}},
  \bibinfo {author} {\bibfnamefont {A.}~\bibnamefont {Avdoshkin}}, \bibinfo
  {author} {\bibfnamefont {T.}~\bibnamefont {Scaffidi}},\ and\ \bibinfo
  {author} {\bibfnamefont {E.}~\bibnamefont {Altman}},\ }\bibfield  {title}
  {\bibinfo {title} {A universal operator growth hypothesis},\ }\href
  {https://doi.org/10.1103/PhysRevX.9.041017} {\bibfield  {journal} {\bibinfo
  {journal} {Phys. Rev. X}\ }\textbf {\bibinfo {volume} {9}},\ \bibinfo {pages}
  {041017} (\bibinfo {year} {2019})}\BibitemShut {NoStop}%
\bibitem [{\citenamefont {Heveling}\ \emph {et~al.}(2022)\citenamefont
  {Heveling}, \citenamefont {Wang},\ and\ \citenamefont
  {Gemmer}}]{HevelingWang}%
  \BibitemOpen
  \bibfield  {author} {\bibinfo {author} {\bibfnamefont {R.}~\bibnamefont
  {Heveling}}, \bibinfo {author} {\bibfnamefont {J.}~\bibnamefont {Wang}},\
  and\ \bibinfo {author} {\bibfnamefont {J.}~\bibnamefont {Gemmer}},\
  }\bibfield  {title} {\bibinfo {title} {Numerically probing the universal
  operator growth hypothesis},\ }\href
  {https://doi.org/10.1103/PhysRevE.106.014152} {\bibfield  {journal} {\bibinfo
   {journal} {Phys. Rev. E}\ }\textbf {\bibinfo {volume} {106}},\ \bibinfo
  {pages} {014152} (\bibinfo {year} {2022})}\BibitemShut {NoStop}%
\bibitem [{\citenamefont {De~Roeck}\ and\ \citenamefont
  {Huveneers}(2014)}]{DeRoeckHuveneers}%
  \BibitemOpen
  \bibfield  {author} {\bibinfo {author} {\bibfnamefont {W.}~\bibnamefont
  {De~Roeck}}\ and\ \bibinfo {author} {\bibfnamefont {F.}~\bibnamefont
  {Huveneers}},\ }\bibfield  {title} {\bibinfo {title} {Scenario for
  delocalization in translation-invariant systems},\ }\href
  {https://doi.org/10.1103/PhysRevB.90.165137} {\bibfield  {journal} {\bibinfo
  {journal} {Phys. Rev. B}\ }\textbf {\bibinfo {volume} {90}},\ \bibinfo
  {pages} {165137} (\bibinfo {year} {2014})}\BibitemShut {NoStop}%
\bibitem [{\citenamefont {De~Roeck}\ and\ \citenamefont
  {Huveneers}(2017)}]{DeRoeckHuveneers2}%
  \BibitemOpen
  \bibfield  {author} {\bibinfo {author} {\bibfnamefont {W.}~\bibnamefont
  {De~Roeck}}\ and\ \bibinfo {author} {\bibfnamefont {F.}~\bibnamefont
  {Huveneers}},\ }\bibfield  {title} {\bibinfo {title} {Stability and
  instability towards delocalization in many-body localization systems},\
  }\href {https://doi.org/10.1103/PhysRevB.95.155129} {\bibfield  {journal}
  {\bibinfo  {journal} {Phys. Rev. B}\ }\textbf {\bibinfo {volume} {95}},\
  \bibinfo {pages} {155129} (\bibinfo {year} {2017})}\BibitemShut {NoStop}%
\bibitem [{\citenamefont {Thiery}\ \emph {et~al.}(2018)\citenamefont {Thiery},
  \citenamefont {Huveneers}, \citenamefont {M\"uller},\ and\ \citenamefont
  {De~Roeck}}]{ThierryHuveneers}%
  \BibitemOpen
  \bibfield  {author} {\bibinfo {author} {\bibfnamefont {T.}~\bibnamefont
  {Thiery}}, \bibinfo {author} {\bibfnamefont {F.}~\bibnamefont {Huveneers}},
  \bibinfo {author} {\bibfnamefont {M.}~\bibnamefont {M\"uller}},\ and\
  \bibinfo {author} {\bibfnamefont {W.}~\bibnamefont {De~Roeck}},\ }\bibfield
  {title} {\bibinfo {title} {Many-body delocalization as a quantum avalanche},\
  }\href {https://doi.org/10.1103/PhysRevLett.121.140601} {\bibfield  {journal}
  {\bibinfo  {journal} {Phys. Rev. Lett.}\ }\textbf {\bibinfo {volume} {121}},\
  \bibinfo {pages} {140601} (\bibinfo {year} {2018})}\BibitemShut {NoStop}%
\bibitem [{\citenamefont {Crowley}\ and\ \citenamefont
  {Chandran}(2020)}]{CrowleyChandran}%
  \BibitemOpen
  \bibfield  {author} {\bibinfo {author} {\bibfnamefont {P.~J.~D.}\
  \bibnamefont {Crowley}}\ and\ \bibinfo {author} {\bibfnamefont
  {A.}~\bibnamefont {Chandran}},\ }\bibfield  {title} {\bibinfo {title}
  {Avalanche induced coexisting localized and thermal regions in disordered
  chains},\ }\href {https://doi.org/10.1103/PhysRevResearch.2.033262}
  {\bibfield  {journal} {\bibinfo  {journal} {Phys. Rev. Res.}\ }\textbf
  {\bibinfo {volume} {2}},\ \bibinfo {pages} {033262} (\bibinfo {year}
  {2020})}\BibitemShut {NoStop}%
\bibitem [{\citenamefont {Morningstar}\ \emph {et~al.}(2022)\citenamefont
  {Morningstar}, \citenamefont {Colmenarez}, \citenamefont {Khemani},
  \citenamefont {Luitz},\ and\ \citenamefont {Huse}}]{MorningstarColmenarez}%
  \BibitemOpen
  \bibfield  {author} {\bibinfo {author} {\bibfnamefont {A.}~\bibnamefont
  {Morningstar}}, \bibinfo {author} {\bibfnamefont {L.}~\bibnamefont
  {Colmenarez}}, \bibinfo {author} {\bibfnamefont {V.}~\bibnamefont {Khemani}},
  \bibinfo {author} {\bibfnamefont {D.}~\bibnamefont {Luitz}},\ and\ \bibinfo
  {author} {\bibfnamefont {D.}~\bibnamefont {Huse}},\ }\bibfield  {title}
  {\bibinfo {title} {Avalanches and many-body resonances in many-body localized
  systems},\ }\href {https://doi.org/10.1103/PhysRevB.105.174205} {\bibfield
  {journal} {\bibinfo  {journal} {Phys. Rev. B}\ }\textbf {\bibinfo {volume}
  {105}},\ \bibinfo {pages} {174205} (\bibinfo {year} {2022})}\BibitemShut
  {NoStop}%
\bibitem [{\citenamefont {Lieb}\ and\ \citenamefont
  {Robinson}(1972)}]{LiebRobinson}%
  \BibitemOpen
  \bibfield  {author} {\bibinfo {author} {\bibfnamefont {E.~H.}\ \bibnamefont
  {Lieb}}\ and\ \bibinfo {author} {\bibfnamefont {D.~W.}\ \bibnamefont
  {Robinson}},\ }\bibfield  {title} {\bibinfo {title} {The finite group
  velocity of quantum spin systems},\ }\href
  {https://doi.org/10.1007/BF01645779} {\bibfield  {journal} {\bibinfo
  {journal} {Commun. Math. Phys.}\ }\textbf {\bibinfo {volume} {28}},\ \bibinfo
  {pages} {251} (\bibinfo {year} {1972})}\BibitemShut {NoStop}%
\bibitem [{\citenamefont {Bravyi}\ \emph {et~al.}(2006)\citenamefont {Bravyi},
  \citenamefont {Hastings},\ and\ \citenamefont {Verstraete}}]{BravyiHastings}%
  \BibitemOpen
  \bibfield  {author} {\bibinfo {author} {\bibfnamefont {S.}~\bibnamefont
  {Bravyi}}, \bibinfo {author} {\bibfnamefont {M.~B.}\ \bibnamefont
  {Hastings}},\ and\ \bibinfo {author} {\bibfnamefont {F.}~\bibnamefont
  {Verstraete}},\ }\bibfield  {title} {\bibinfo {title} {Lieb-robinson bounds
  and the generation of correlations and topological quantum order},\ }\href
  {https://doi.org/10.1103/PhysRevLett.97.050401} {\bibfield  {journal}
  {\bibinfo  {journal} {Phys. Rev. Lett.}\ }\textbf {\bibinfo {volume} {97}},\
  \bibinfo {pages} {050401} (\bibinfo {year} {2006})}\BibitemShut {NoStop}%
\bibitem [{\citenamefont {Imbrie}(2016{\natexlab{b}})}]{Imbrie2016}%
  \BibitemOpen
  \bibfield  {author} {\bibinfo {author} {\bibfnamefont {J.~Z.}\ \bibnamefont
  {Imbrie}},\ }\bibfield  {title} {\bibinfo {title} {Diagonalization and
  many-body localization for a disordered quantum spin chain},\ }\href
  {https://doi.org/10.1103/PhysRevLett.117.027201} {\bibfield  {journal}
  {\bibinfo  {journal} {Phys. Rev. Lett.}\ }\textbf {\bibinfo {volume} {117}},\
  \bibinfo {pages} {027201} (\bibinfo {year} {2016}{\natexlab{b}})}\BibitemShut
  {NoStop}%
\bibitem [{\citenamefont {{De Roeck}}\ \emph {et~al.}(2024)\citenamefont {{De
  Roeck}}, \citenamefont {Giacomin}, \citenamefont {Huveneers},\ and\
  \citenamefont {Prosniak}}]{deRoeckGiacomin}%
  \BibitemOpen
  \bibfield  {author} {\bibinfo {author} {\bibfnamefont {W.}~\bibnamefont {{De
  Roeck}}}, \bibinfo {author} {\bibfnamefont {L.}~\bibnamefont {Giacomin}},
  \bibinfo {author} {\bibfnamefont {F.}~\bibnamefont {Huveneers}},\ and\
  \bibinfo {author} {\bibfnamefont {O.}~\bibnamefont {Prosniak}},\ }\href
  {https://arxiv.org/abs/2408.04338} {\bibinfo {title} {Absence of normal heat
  conduction in strongly disordered interacting quantum chains}} (\bibinfo
  {year} {2024}),\ \Eprint {https://arxiv.org/abs/2408.04338} {arXiv:2408.04338
  [math-ph]} \BibitemShut {NoStop}%
\bibitem [{\citenamefont {Mazur}(1969)}]{Mazur}%
  \BibitemOpen
  \bibfield  {author} {\bibinfo {author} {\bibfnamefont {P.}~\bibnamefont
  {Mazur}},\ }\bibfield  {title} {\bibinfo {title} {Non-ergodicity of phase
  functions in certain systems},\ }\href
  {https://doi.org/10.1016/0031-8914(69)90185-2} {\bibfield  {journal}
  {\bibinfo  {journal} {Physica}\ }\textbf {\bibinfo {volume} {43}},\ \bibinfo
  {pages} {533} (\bibinfo {year} {1969})}\BibitemShut {NoStop}%
\bibitem [{\citenamefont {Prosen}(2011)}]{Prosen}%
  \BibitemOpen
  \bibfield  {author} {\bibinfo {author} {\bibfnamefont {T.}~\bibnamefont
  {Prosen}},\ }\bibfield  {title} {\bibinfo {title} {Open $xxz$ spin chain:
  Nonequilibrium steady state and a strict bound on ballistic transport},\
  }\href {https://doi.org/10.1103/PhysRevLett.106.217206} {\bibfield  {journal}
  {\bibinfo  {journal} {Phys. Rev. Lett.}\ }\textbf {\bibinfo {volume} {106}},\
  \bibinfo {pages} {217206} (\bibinfo {year} {2011})}\BibitemShut {NoStop}%
\bibitem [{\citenamefont {Sirker}\ \emph {et~al.}(2009)\citenamefont {Sirker},
  \citenamefont {Pereira},\ and\ \citenamefont {Affleck}}]{SirkerPereira}%
  \BibitemOpen
  \bibfield  {author} {\bibinfo {author} {\bibfnamefont {J.}~\bibnamefont
  {Sirker}}, \bibinfo {author} {\bibfnamefont {R.~G.}\ \bibnamefont
  {Pereira}},\ and\ \bibinfo {author} {\bibfnamefont {I.}~\bibnamefont
  {Affleck}},\ }\bibfield  {title} {\bibinfo {title} {Diffusion and ballistic
  transport in one-dimensional quantum systems},\ }\href
  {https://doi.org/10.1103/PhysRevLett.103.216602} {\bibfield  {journal}
  {\bibinfo  {journal} {Phys. Rev. Lett.}\ }\textbf {\bibinfo {volume} {103}},\
  \bibinfo {pages} {216602} (\bibinfo {year} {2009})}\BibitemShut {NoStop}%
\bibitem [{\citenamefont {Sirker}(2020)}]{SirkerLectureNotes}%
  \BibitemOpen
  \bibfield  {author} {\bibinfo {author} {\bibfnamefont {J.}~\bibnamefont
  {Sirker}},\ }\bibfield  {title} {\bibinfo {title} {{Transport in
  one-dimensional integrable quantum systems}},\ }\href
  {https://doi.org/10.21468/SciPostPhysLectNotes.17} {\bibfield  {journal}
  {\bibinfo  {journal} {SciPost Phys. Lect. Notes}\ ,\ \bibinfo {pages} {17}}
  (\bibinfo {year} {2020})}\BibitemShut {NoStop}%
\bibitem [{\citenamefont {Pereira}\ \emph {et~al.}(2014)\citenamefont
  {Pereira}, \citenamefont {Pasquier}, \citenamefont {Sirker},\ and\
  \citenamefont {Affleck}}]{PereiraPasquier}%
  \BibitemOpen
  \bibfield  {author} {\bibinfo {author} {\bibfnamefont {R.~G.}\ \bibnamefont
  {Pereira}}, \bibinfo {author} {\bibfnamefont {V.}~\bibnamefont {Pasquier}},
  \bibinfo {author} {\bibfnamefont {J.}~\bibnamefont {Sirker}},\ and\ \bibinfo
  {author} {\bibfnamefont {I.}~\bibnamefont {Affleck}},\ }\bibfield  {title}
  {\bibinfo {title} {Exactly conserved quasilocal operators for the xxz spin
  chain},\ }\href {https://doi.org/10.1088/1742-5468/2014/09/P09037} {\bibfield
   {journal} {\bibinfo  {journal} {J. Stat. Mech.}\ ,\ \bibinfo {pages}
  {P09037}} (\bibinfo {year} {2014})}\BibitemShut {NoStop}%
\end{thebibliography}
%

\end{document}